\documentclass[sn-mathphys-num]{sn-jnl}

\usepackage{graphicx}%
\usepackage{multirow}%
\usepackage{amsmath,amssymb,amsfonts}%
\usepackage{amsthm}%
\usepackage{mathrsfs}%
\usepackage[title]{appendix}%
\usepackage{xcolor}%
\usepackage{textcomp}%
\usepackage{manyfoot}%
\usepackage{booktabs}%
\usepackage{algorithm}%
\usepackage{algorithmicx}%
\usepackage{algpseudocode}%
\usepackage{listings}%
\usepackage{natbib}
\usepackage{comment}

\newcommand{\mb}[1]{\boldsymbol{\mathbf{#1}}}

\theoremstyle{thmstyleone}%
\theoremstyle{thmstyletwo}%

\theoremstyle{thmstylethree}%

\begin{document}
	
	\title{TransST: Transfer Learning Embedded Spatial Factor Modeling of Spatial Transcriptomics Data}
	
	
	\author[1]{\fnm{Shuo Shuo} \sur{Liu}}\email{shuoliu.academic@gmail.com}
	
	\author[1]{\fnm{Shikun} \sur{Wang}}\email{sw3763@cumc.columbia.edu}

 \author[1]{\fnm{Yuxuan} \sur{Chen}}\email{yc4018@cumc.columbia.edu}
	
	\author[2]{\fnm{Anil K.} \sur{Rustgi}}\email{akr2164@cumc.columbia.edu}
	
	\author[3]{\fnm{Ming} \sur{Yuan}}\email{ming.yuan@columbia.edu}
	
	\author*[1]{\fnm{Jianhua} \sur{Hu}}\email{jh3992@cumc.columbia.edu}
	
	\affil[1]{\orgdiv{Department of Biostatistics}, \orgname{Columbia University}, \orgaddress{\postcode{10032}, \state{NY}, \country{United States}}}

	\affil[2]{\orgdiv{Department of Medicine}, \orgname{Columbia University}, \orgaddress{\postcode{10027}, \state{NY}, \country{United States}}}
	
	\affil[3]{\orgdiv{Department of Statistics}, \orgname{Columbia University}, \orgaddress{\postcode{10027}, \state{NY}, \country{United States}}}

	
	\abstract{
    \textbf{Background:} Spatial transcriptomics have emerged as a powerful tool in biomedical research because of its ability to capture both the spatial contexts and abundance of the complete RNA transcript profile in organs of interest. 
		However, limitations of the technology such as the relatively low resolution and comparatively insufficient sequencing depth make it difficult to reliably extract real biological signals from these data. To alleviate this challenge, we propose a novel transfer learning framework, referred to as TransST, to adaptively leverage the cell-labeled information from external sources in inferring cell-level heterogeneity of a target spatial transcriptomics data. 

        \textbf{Results:} Applications in several real studies as well as a number of simulation settings show that our approach significantly improves existing techniques. 
		For example, in the breast cancer study, TransST successfully identifies five biologically meaningful cell clusters, including the two subgroups of cancer in situ and invasive cancer; in addition, only TransST is able to separate the adipose tissues from the connective issues among all the studied methods. 
        
        \textbf{Conclusions:} In summary, the proposed method TransST is both effective and robust in identifying cell subclusters and detecting corresponding driving biomarkers in spatial transcriptomics data.}

\keywords{Clustering, factor model, Markov random field, Spatial transcriptomics, Transfer learning}



\maketitle

\section{Introduction}
\label{sec: intro}
Recent advancements in high-throughput genomics technologies have revolutionized our understanding of cellular heterogeneity and spatial organization within tissues. Single-cell RNA sequencing (scRNA-seq) has been a pivotal tool for unraveling the landscape of gene expression at the single-cell level, providing unprecedented insight into the molecular diversity of complex tissues. 
While scRNA-seq enables a high-throughput examination of cell transcriptomes, the spatial context is forfeited during tissue processing. 
In contrast, methods like immunohistochemistry (IHC) and in situ hybridization offer elevated spatial resolution but frequently demand pre-selection of targets, rendering them less suitable for high-throughput exploratory analyses.
Spatial transcriptomics has emerged as a powerful approach to bridge this gap by directly preserving the spatial coordinates of gene expression patterns within tissue sections. This technology facilitates the exploration of the spatial organization of cells and the identification of spatially coordinated gene expression, offering a holistic perspective on the complex interplay of cells within their native niches.
Spatial transcriptomics technology has been widely applied in many biomedical studies, such as the human postmortem DLPFC tissue \citep{yang2022sc,lin2022scjoint}, area identification of mouse embryos \citep{lohoff2022integration}, and region identification of squamous cell carcinoma \citep{ji2020multimodal}.

In this article, we aim to delineate the cell heterogeneity via clustering cells or spots, which are cell mixtures, into meaningful subtypes of spatial transcriptomics (ST) data, boosted further by borrowing available external information.
Clustering methods based on either the high-dimensional scRNA-seq data or the high-dimensional spatial transcriptomics data have been extensively studied in the literature.
There are mainly two clustering approaches---tandem or joint.
For the tandem approaches, dimension-reduction techniques are typically employed to achieve a low-dimensional representation. 
This preliminary step precedes the subsequent (spatial) clustering analysis.
Upon attaining a low-dimensional representation through dimension reduction, various clustering methods are popularly employed in scRNA-seq analysis, including kmeans and the Gaussian mixture model (GMM). Spatial transcriptomics analysis utilizes several spatial clustering methods, such as the graph convolutional network (GCN)-based approach SpaGCN \citep{hu2021spagcn}, Giotto \citep{dries2021giotto}, BayesSpace \citep{zhao2021spatial}, and SC-MEB \citep{yang2022sc}. 
The joint approach estimates the low-dimensional representations and the latent cell clustering membership labels simultaneously, such as FKM \citep{markos2019beyond} and ORCLUS \citep{aggarwal2000finding} for non-spatial clustering, and DR.SC \citep{liu2022joint} for spatial clustering.
However, all above-mentioned ST analysis methods use the data generated from the current ST studies alone, while not taking advantage of data generated from other existing studies relevant to the concerning disease type, including scRNA-seq data and ST data.

Our goal is to leverage the useful information of existing studies (scRNA-seq data or ST data) to learn a target/new study of the similar diseases. 
Transfer learning, a paradigm within machine learning, allows the transfer of knowledge gained from one domain to another, overcoming the limitations posed by insufficient or not labeled data in the target domain. 
In the context of spatial transcriptomics, transfer learning can be leveraged to exploit the rich information encoded in scRNA-seq data to enhance the clustering and interpretation of spatial expression profiles. By learning relevant features from the source data and adapting them to the target domain, transfer learning holds the potential to improve the accuracy and robustness of spatial transcriptomics analysis.
Transfer learning has been studied 
in the clustering of scRNA-seq data in the literature.
For example, non-negative matrix factorization with transfer learning was shown to improve the clustering of scRNA-seq data by utilizing prior reference knowledge \citep{mieth2019using}.
A transfer learning model under an iterative non-negative matrix factorization framework is further studied \citep{peng2021integration}, in which the information is transferred from low-rank matrix of more informative data to data of lower quality.
Transfer learning in neural networks also has been investigated in the literature.
For example, a transfer learning algorithm borrows the information from supervised cell type classification algorithms, which improves clustering of the scRNA-seq data by building up neural networks \citep{hu2020iterative}. 
A transfer learning model is developed to integrate atlas-scale scRNA-seq and ATAT-seq data in a semi-supervised neural network framework \citep{lin2022scjoint}.
All the above-mentioned methods focus on transfer learning without utilization of the essential spatial characteristics from the spatial transcriptomics data.

In this article, we aim at proposing a transfer learning embedded systematic statistical framework to improve spatial transcriptomics data analysis using rich information extracted from external scRNA-seq or ST data, which is abbreviated as {\bf TransST}.
Our proposed integrative approach differentiates from the conventional methods, which focuses on analyzing a target spatial transcriptomics data alone. 
This novel methodology integrates both source (e.g., scRNA-seq) and target (e.g., spatial transcriptomics) data, introducing a paradigm shift in the cell subtyping and biomarker detection process. 
While the traditional methods focus solely on the spatial domain, our integrative model leverages the cell clustering membership labeled information from scRNA-seq, providing a unique advantage for ST data analysis. 
Note that our method is applicable across different platforms, specifically sequencing-based and image-based spatial transcriptomics data.

We conduct extensive simulation studies and investigate four real applications to demonstrate the promising performance of the proposed TransST. Here are the examples of some biologically interesting findings in our real data studies. In the HER2-positive breast cancer study, TransST successfully identifies five biologically meaningful cell clusters, including the two subgroups of cancer in situ and invasive cancer; in addition, only TransST is able to separate the adipose tissues from the connective issues among all the studied methods. In the cSCC study, we identify a new cell cluster which appears to be highly associated with immune processes, represented by the corresponding top driving genes DSCI, HOPX, and several others.

To summarize, the advantages of our integrative approach arise in several key aspects. 
First, incorporating the information of scRNA-seq data enhances the robustness of cell clustering, offering a more accurate identification and interpretation of cellular heterogeneity by capturing nuanced gene expression patterns.  
Second, we tackle the challenges associated with the integration of these distinct data types and propose an adaptive approach to estimate the target model parameters.
Adaptively transferring information from the source data to the target data provides effective and powerful spatial transcriptomics data analysis.

\section{Methods}
\label{sec: method}

\textbf{Notations:} We use ``spot" to denote a location that corresponds to either a cell or a cell mixture in spatial transcriptomics data, depending on the specific technology.  
Let $\mb X_{src}\in\mathbb R^{n_1\times p}$ and $\mb X_{tgt}\in\mathbb R^{n_0\times p}$ be the $p$-dimensional gene expression matrices from the source data with $n_1$ cells/spots and the target data with $n_0$ cells/spots, respectively. Denote $\mb U$ and $\mb V$ as the corresponding low-dimensional representations of $\mb X_{src}$ and $\mb X_{tgt}$.
We use the lowercase letters with subscript $i$ to denote the $p$-dimensional vector for the $i$-th spot.  
For example, $\mb x_{src,i},\mb u_i,\mb x_{tgt,i},\mb v_i$ represent the $i$-th row of $X_{src},\mb U,X_{tgt}$ and $\mb V$, respectively.
For simplicity, we use the same notations $z_i$, $\mb\mu_k$, and $\mb\Sigma_k$ for both the low-dimensional source and target data.
Particularly, we use the same $\mb\mu_k$ and $\mb\Sigma_k$ for simplicity but they differ in Step 1 and Step 3.
As described in the Result section, multiple external samples can be combined into a single source data with proper gene expression normalization.

\begin{figure*}
	\centering
	\includegraphics[width=0.9\linewidth,keepaspectratio]{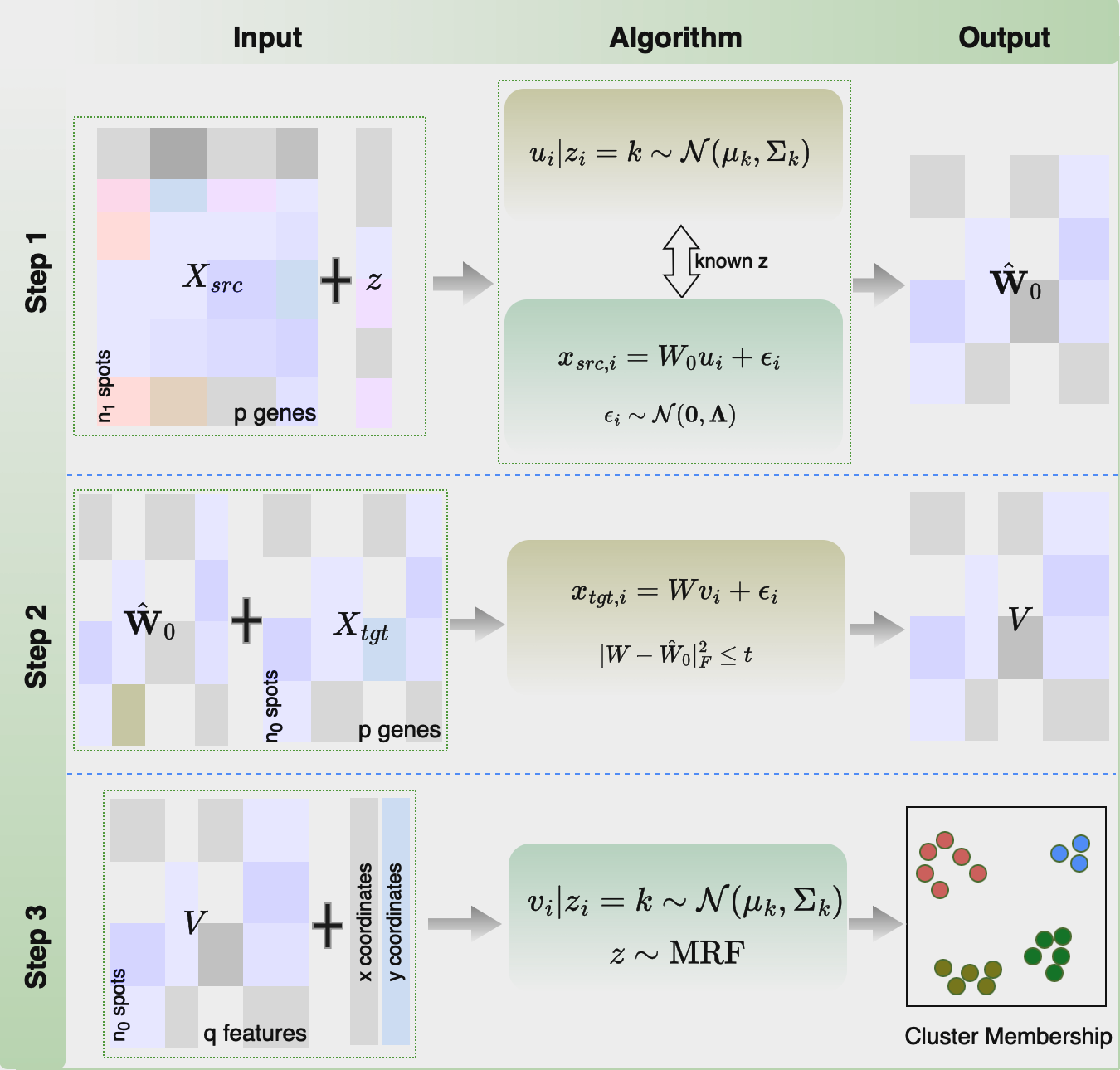}
	\caption{Overview of TransST. TransST involves three steps. Step 1 learns the weight matrix from the source data; Step 2 adaptively updates the weight matrix of the target data by leveraging information transferred from the source data; Step 3 models the low-dimensional representation for downstream analysis, such as cluster analysis and identification of differentially expressed genes.}
	\label{fig: flow}
\end{figure*}

Existing methods either focus on analyzing scRNA-seq data in a single study, or consider information transferring between different scRNA-seq data sets which have some related disease mechanisms. Distinctive from those, our method intends to build up a more general framework for transferring useful knowledge learned from biologically related scRNA-seq or spatial transcriptomics data of individual samples or different studies into the target spatial transcriptomics data set, with the ultimate goal of improving cell heterogeneity identification and gene biomarker detection. 
Figure \ref{fig: flow} shows a flowchart of our proposed TransST.
More specifically, TransST takes the preprocessed gene expression matrices, the cell types from the source data, and the spatial coordinates in the target spatial transcriptomics data as the input.
It factorizes the high-dimensional input into a low-dimensional representation with a learned loading matrix $\hat{\mb W}$, guided by the known cell types $\mb z$ from the source data.
In Step 2, we estimate the low-dimensional representation of the target data based on a factor model, while reserving the learned information of the loading matrix from Step 1.
This step mimics the recent developments about semi-supervised learning and domain adaptation in deep learning but in a parametric manner.
In Step 3, the spatial clustering step, we employ a Markov random field (MRF) to account for spatial smoothness within the spot neighborhoods in the cluster label space.

\textbf{Step 1. pLDR: Probabilistic Linear Dimension Reduction}

Since the source data $\mb X_{src}$ is typically high-dimensional, we consider the following probabilistic linear dimension reduction model (pLDR).
\begin{equation*}
	\begin{split}
		\mb x_{src,i}&=\mb W_0\mb u_i+\mb\epsilon_i, \mb\epsilon_i\sim\mathcal{N}(\mb 0,\mb\Lambda),\\
		\mb u_i|z_i=k&\sim\mathcal{N}(\mb\mu_k,\mb\Sigma_k).
	\end{split}
\end{equation*}
We project the $p$-dimensional gene expression $\mb x_{src, i}$ into a $q$-dimensional representation $\mb u_i$ with a factor loading matrix $\mb W_0\in\mathbb{R}^{p\times q}$; the residual $\mb \epsilon_i$ is assumed to follow a zero mean multivariate normal distribution with variance-covariance matrix
$\mb\Lambda$, a diagonal matrix for residual variance.
Denote $z_i$ as the cluster label for spot $i$.
Given $z_i=k$ that spot $i$ belongs to the $k$-th cluster, $\mb u_i$ follows a multivariate normal distribution with mean
$\mb\mu_k\in\mathbb{R}^{q\times 1}$ and covariance matrix $\mb\Sigma_k\in\mathbb{R}^{q\times q}$.
Note that the cluster labels (cell types) are known in the source data for Step 1, so pLDR indeed learns $\mb W_0$ in a supervised learning setting.

\textbf{Step 2. Adaptive transfer learning}

Next, we utilize the estimated $\hat{\mb W}_0$ learned in Step 1 to reduce the dimension of $\mb x_{tgt,i}$. Specifically, we estimate the low-dimensional representation $\mb v_i$ and latent loading matrix $\mb W$ by minimizing
\begin{equation*}
	\label{eq: what}
	\sum_{i=1}^{n_0}\|\mb x_{tgt,i}-\mb W\mb v_i\|_2^2+\lambda \|\mb W-\hat{\mb W}_0\|_F^2.
\end{equation*}
This step learns $\mb v_i$ by penalizing $\mb W$ towards $\hat{\mb W}_0$ using a Frobenius norm.
The tuning parameter $\lambda$ is chosen in a similar manner to cross-validation.
More specifically, we evenly split the target data into two folds.
Within each fold, we first calculate the low-dimensional representation and the loading matrix.
Next, we use the loading matrix from one fold to reconstruct $\mb x_{tgt}$ and record the reconstruction error.
For a given set of $\lambda$, we choose the one that minimizes the reconstruction error.

\textbf{Step 3. spGMM: spatial Gaussian mixture model}

With the estimated low-dimensional representation $\hat{\mb v}_i$, we consider a spatial Gaussian mixture model for clustering, i.e.,
\begin{equation*}
	\begin{split}
		f(\mb v_i|z_i=k)&\sim\mathcal{N}(\mb\mu_k,\mb\Sigma_k),\\
		f(\mb z) &\propto\exp\left\{-\frac12\sum_{i=1}^{n_0}\sum_{i'\in N_i}\beta \left(1-I\{z_i=z_{i'}\}\right)\right\}.
	\end{split}
\end{equation*}
Given $z_i=k$, $\mb v_i$ follows a multivariate normal distribution with mean
$\mb\mu_k\in\mathbb{R}^{q\times 1}$ and covariance matrix $\mb\Sigma_k\in\mathbb{R}^{q\times q}$. 
The smoothing parameter $\beta$ controls the signal strength from the spatial coordinates \citep{liu2022joint}. 
We choose $\beta\in[0,1]$ that maximizes the likelihood by line search.
Particularly, $\beta=0$ implies spatial coordinates are redundant and spGMM returns to the original GMM.
We defer to supplementary material on the selection of number of clusters $K$.
$I\{z_i=z_{i'}\}$ is an indicator function and $N_i$ is the neighborhood of the $i$-th spot, constructed by the spatial coordinates.

Different from the bisection method used in DR.SC and SC-MEB, we use the $k$-nearest neighbor (knn) algorithm to define neighbors. 
In all experiments, we use 5-nearest neighbor algorithm.
By conducting extensive experiments, we find that knn works better than the bisection method used in SC-MEB and DR.SC.

\section{Results}
\label{sec: res}

To evaluate the performances of TransST, we compare it with several popular methods, including the ones based on dimension reduction and Gaussian mixtures, which enjoy the benefits of interpretability, computational efficiency, and modeling simplicities.
These methods include DR.SC \citep{liu2022joint}, GMM, kmeans, SC-MEB \citep{yang2022sc}, Seurat-v5 \citep{hao2023dictionary}, Louvain \citep{blondel2008fast}, and Leiden \citep{traag2019louvain}. 
We calculate the adjusted rand index (ARI) to quantify the clustering performance.
To further demonstrate the clustering accuracy of TransST, we also compare it to GraphST \citep{long2023spatially} (a graph self-supervised contrastive learning method), knn, and random forest (RF).
We also compare the computational time of different methods.
We refer the interested readers to the supplementary materials for more details.

We demonstrate the superior performance of TransST through extensive simulation studies and four real studies.
In the four real studies, we compare ARIs of various methods whenever the annotated cluster labels are available.
First, we investigate TransST in the breast cancer data and discover cell types based on spatial mapping of ST data.
Next, we demonstrate that TransST is also an effective tool to help understand the organization and development of organs and tissues, such as identifying brain layers and distinguishing areas of the mouse embryo in the second and third real studies, respectively.
The available annotated cell subgroup structures, which are assumed to be the truth, in the first three studies provide strong evidence of the reliability and effectiveness of TransST. 
The result also demonstrates its robust generalizability across diverse types of datasets. 
Furthermore, it also shows the promising performance in terms of important pattern recognition and biomarker detection in the fourth study of squamous cell carcinoma data.
All studies in this section use the function \textit{FindAllMarkers} from package Seurat-V5 for identifying DEGs.

\subsection{Validation using simulated data}
\label{sec: simu}

We conduct comprehensive simulation studies to evaluate the performance of TransST as well as several popular methods for comparisons. 
We provide simulation details in the Method section.
Briefly, we first simulate the cell types by the Potts model. 
Then we generate the low-dimensional representation by a multivariate normal distribution and multiply it by a factor loading matrix, which results in the gene expression matrix.

\begin{figure*}[hbt!]
	\centering
	\includegraphics[width=\linewidth,keepaspectratio]{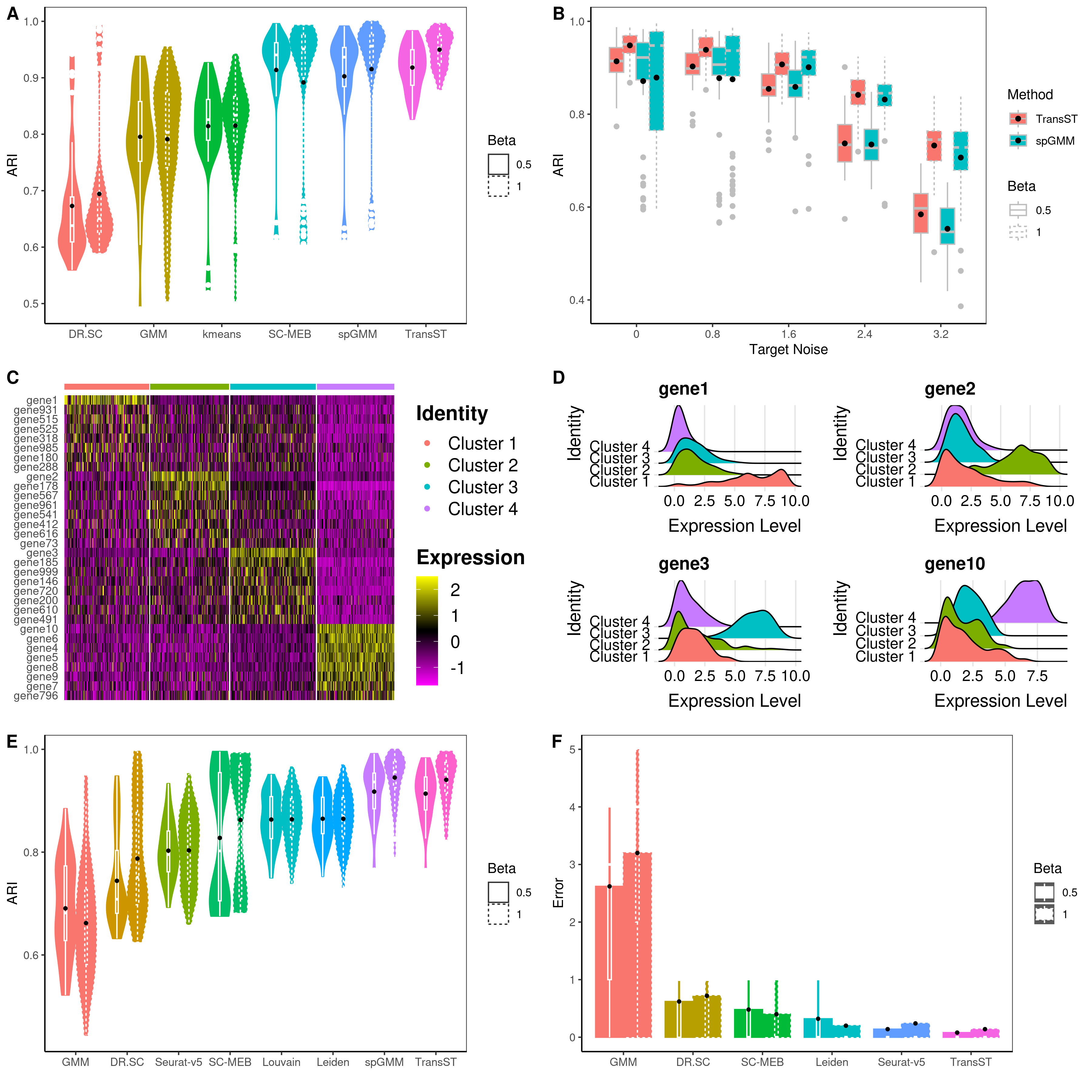}
	\caption{Comparison of TransST with multiple methods across various simulation settings. 
		Black dots represent the averaged values. 
		A: Comparison of the clustering performances among various methods when $\beta=0.5$ (solid line) and $\beta=1$ (dotted line), using the true number of clusters;
		B: Transferring power of TransST with different levels of noise in the target data;
		C: Heatmap of differentially expressed genes for each cell type identified by TransST with $\beta=1$;
		D: Visualization the distribution of gene expression across different cell clusters by ridge plot;
		E: Comparison of the clustering performances among various methods that can estimate the number of clusters;
		F: Absolute error $|\hat K-K|$ of estimating the number of clusters by various methods.}
	\label{fig: simu}
\end{figure*}

Figure \ref{fig: simu} presents the simulation results, including the Adjusted Rand Index (ARI), selection of number of clusters, and identification of differentially expressed genes (DEGs).
In the case of tandem (low-dimensional) methods such as kmeans, GMM, Louvain, Leiden, SC-MEB, and spGMM, we implement principal component analysis (PCA) as the input.
Figure \ref{fig: simu}A shows the effects of the spatial signal strength $\beta$ when the number of clusters is known.
First, we notice that ARIs of DR.SC, SC-MEB, spGMM, and TransST increase as $\beta$ increases, which implies the benefit of clustering from a stronger spatial signal.
Second, by comparing ARIs of GMM with those of spGMM, we conclude that the inclusion of spatial information significantly improves the clustering performance in our simulation setting.
Third, we notice that transferring information from the source data improves the clustering performance in the target data.
This evidence can be collected by comparing TransST to spGMM and SC-MEB.
In Figure \ref{fig: simu}B, we demonstrate the benefits of the transfer learning method TransST over the non-transfer version spGMM under different noise levels in the target data.
To achieve this, we introduce a noise matrix to $\mb X_{tgt}$,  with elements consisting of $e$ multiplied by a standard normal random variable.
Overall, we observe an increase in ARIs in both spGMM and TransST as the strength of the spatial signal intensifies.
As the noise level increases, the signals become much less clear with worsened masking of noises and the target data becomes harder to be clustered. 
Therefore, the improvement after borrowing information from the source data becomes marginal.
In Figures \ref{fig: simu}C, we show the identified DEG by our TransST.
Recall that genes 1, 2, 3, 4--10 uniquely determine clusters 1--4, respectively.
The heatmap shows that the TransST approach can identify DEGs for each cluster.
In Figures \ref{fig: simu}D, the ridge plot shows the distributions of top-selected genes in each cluster.
For example, gene 1 is highly expressed in Cluster 1 whereas it does not express at all in the remaining clusters.
In addition, Figures \ref{fig: simu}E and F show the simulation results when the number of clusters is unknown.
Our TransST accurately estimates the true number of clusters and maintains the highest ARIs.

To further demonstrate the efficiency in terms of computational time, we consider four scenarios with the same setting as Section \ref{sec: simu} and $p=200$: (1) s1: $n_0=10000$, $n_1=2500$; (2) s2: $n_0=20000$, $n_1=2500$; (3) s3: $n_0=10000$, $n_1=5000$; (4) s4: $n_0=10000$, $n_1=10000$.
In all four scenarios, the computational time of the proposed TransST remains under one minute, which demonstrates the scalability for tens of thousands of cells or spots. The computational time of all methods except TransST remains about the same since they do not borrow information from source data.
As $n_0$ increases from 10000 to 20000 (s1 to s2) fixing other parameters, the computational time of methods including DR.SC, SC-MEB, Seurat, spGMM, and TransST increases. 
For TransST, an increasing $n_1$ does not increase the computational time significantly.

\begin{figure}
	\centering
\includegraphics[width=\linewidth,keepaspectratio]{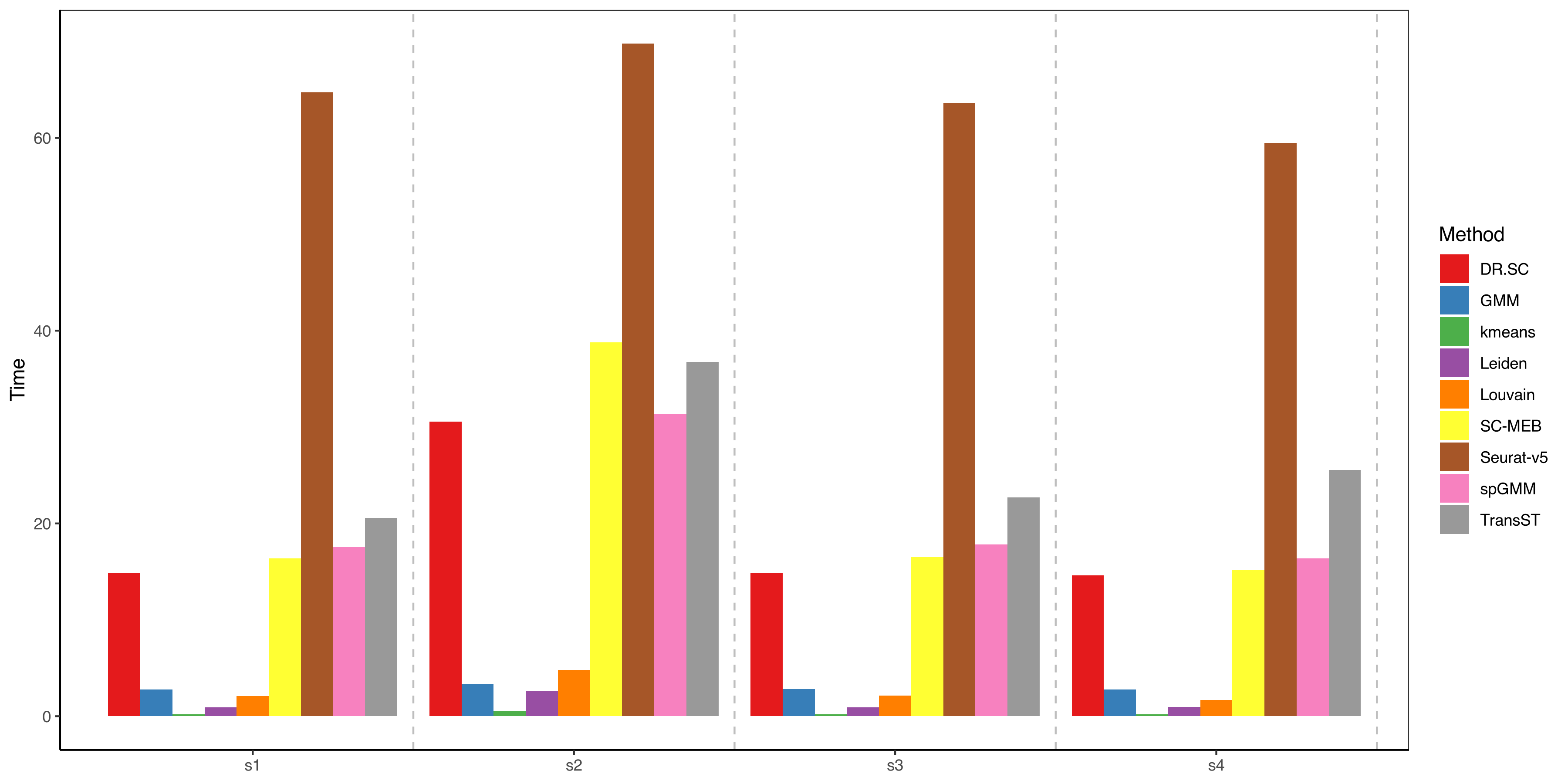}
	\caption{Computational time (in seconds) of various methods in different settings.}
\end{figure}

\subsection{TranST enables accurate spatial mapping of scRNA-seq data in human HER2-positive tumors}

\begin{figure*}[hbt!]
	\centering
	\includegraphics[width=\linewidth,keepaspectratio]{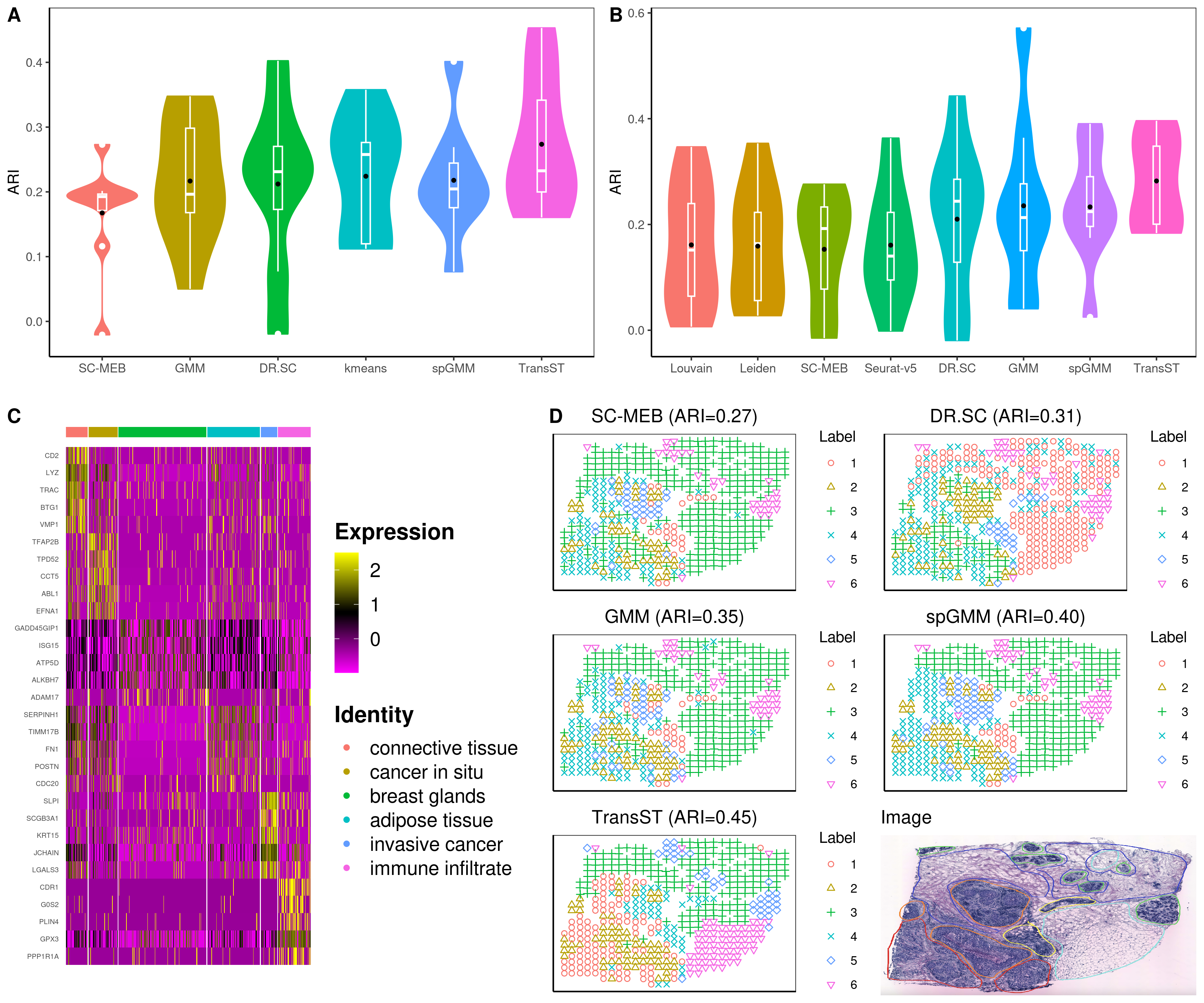}
	\caption{TranST enables accurate spatial mapping of scRNA-seq data in human HER2-positive tumors. 
		A: Comparison of the clustering performance among various methods, using the true number of clusters;
		B: Comparison of methods that can estimate the number of clusters;
		C: Heatmap of the top 5 differentially expressed genes for each cell type in Sample H1, identified by TransST. Each cluster is annotated based on its association with morphological regions;
		D: Spatial heatmaps with estimated labels by various methods for Sample H1.
		Morphological regions in the Image are annotated by a pathologist into six distinct categories: adipose tissue (cyan), breast glands (green), cancer in situ (orange), connective tissue (blue), immune infiltrate (yellow), and invasive cancer (red).}
	\label{fig: H1}
\end{figure*}

HER2-positive tumors in breast cancer refer to a specific subtype of breast cancer that overexpresses the human epidermal growth factor receptor 2 (HER2) protein. HER2 is a protein that plays a role in the regulation of cell growth and division. When breast cancer cells have an increased number of copies of the HER2 gene or produce too much HER2 protein, it can lead to uncontrolled cell growth and a more aggressive form of breast cancer \citep{andersson2021spatial}.
Therefore, investigation into the molecular mechanisms governing the progression of breast tumors has become an important area.

We download the HER2 positive breast tumors from \url{https://github.com/almaan/her2st}, generated with spatial coordinates.
We take samples A1, B1, C1, D1, E1, F1, G2, and H1 for analysis since only the metadata of these samples are publicly available.
A pathologist examines and annotates one section from each tumor, relying on the morphology depicted in the associated Hematoxylin and Eosin (HE) image. 
Regions are categorized as follows: in situ cancer, invasive cancer, adipose tissue, immune infiltrate, or connective tissue.
More details about these samples can be obtained from \cite{andersson2021spatial}. 
For our TransST, genes in the source data are chosen to be the same as those in the target data.
All methods use the same lower dimension $q=15$ and the tandem methods use the PCA results as the inputs.
For our TransST, we treat one sample as the target data and the remaining samples as the source data each time. 
All the other methods in comparison analyze the target data only, without using the information from the remaining samples.

Figures \ref{fig: H1}A and \ref{fig: H1}B illustrate the clustering results when the number of clusters $K$ is known and unknown, respectively.
Our TransST outperforms the others in both scenarios with little sacrifice of ARI when $K$ is unknown.
Figure \ref{fig: H1}C shows the top 5 differentially expressed genes across the seven identified cell types.
In fact, many identified genes have been discussed in the biological literature, which supports the meaningfulness of our obtained cell types. 
For example, gene CD2 is found to have a positive correlation with most
immune infiltrating cells and CD2 is associated with immune responses  \citep{wu2020comprehensive, chen2021cd2};
The tumoral expression of lysozyme (LYZ) correlates with breast cancer lesions that have a favorable prognosis \citep{vizoso2001lysozyme, serra2002expression};
TRAC (T Cell Receptor Alpha Constant) is a Protein Coding gene and
T cell infiltration significantly influences patient survival across all breast cancer subtypes \citep{tietscher2023comprehensive};
BTG1 (BTG Anti-Proliferation Factor 1) is a well-studied member of an anti-proliferative gene family in biology \citep{woo2019promoter};
High expression of the vacuole membrane protein 1 (VMP1) serves as a potential indicator of poor prognosis in HER2-positive breast cancer \citep{amirfallah2019high}.
Based on these observations, we infer the first cluster as immune infiltrate.
TFAP2B (AP-2$\beta$) controls cell proliferation in lobular carcinoma in situ and AP-2$\beta$-negative shows more often in cases of ductal carcinoma in situ (DCIS) \citep{raap2018lobular};
The in situ TPD52 expression pattern is also characterized in the literature \citep{byrne1995screening, balleine2000hd52}.
Based on these, we identify the second cluster as cancer in situ.
SLPI is downregulated in breast cancer than in normal tissues \citep{xie2019expression};
Results reveal that SCGB3A1 promoter methylation is notably elevated in the tumor group and correlated significantly with various clinicopathologic features in breast cancer \citep{nomair2021scgb3a1};
Low expression of KRT15 is significantly linked to a poor prognosis in patients with BRCA, suggesting that KRT15 may play a crucial role in the progression of breast invasive carcinoma \citep{zhong2021low}.
Thus, we infer the fifth cluster as invasive cancer.
PLIN4 may be involved in packaging triacylglycerol into adipocytes \citep{sirois2019unique}, from which we infer the sixth cluster as adipose tissue.
Spatial heatmap in Figure \ref{fig: H1}D further demonstrates the performance of TransST.
For example, TransST is able to separate the adipose tissue from the connective tissue (cyan and blue in the Image; Label 3 and Label 6 from TransST) while SC-MEB, DR.SC, GMM, and spGMM mix them together.

\subsection{TransST enables accurate identification of brain layers of the DLPFC dataset}

\begin{figure*}[hbt!]
	\centering
	\includegraphics[width=\linewidth,keepaspectratio]{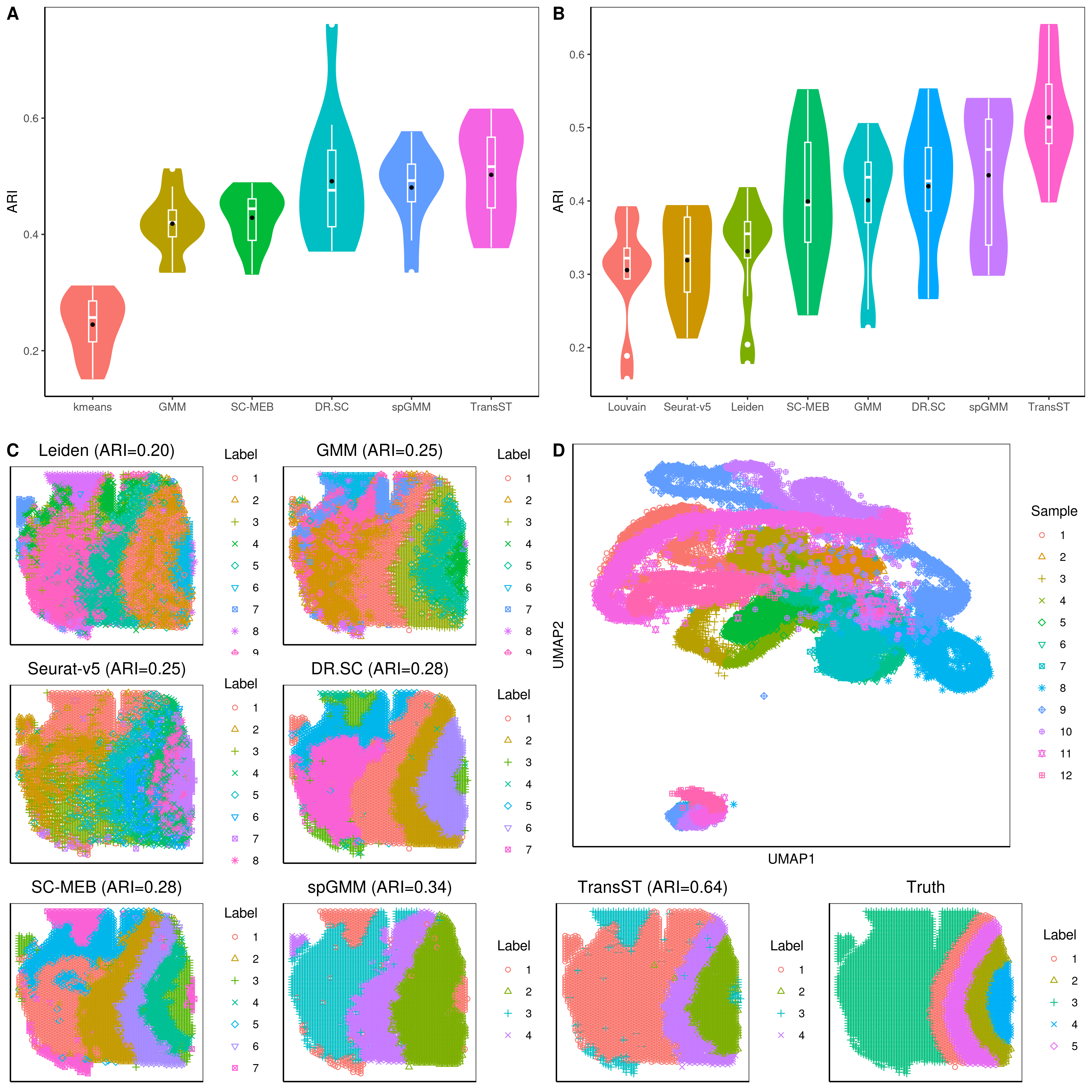}
	\caption{TransST enables accurate identification of brain layers of the DLPFC dataset. 
		A: Comparison of the clustering performance among various methods, using the true number of clusters;
		B: Comparison of methods that can estimate the number of clusters;
		C: Spatial heatmaps with estimated labels by various methods for Sample 151669 ($K$ is unknown);
		D: UMAP plots for TransST with colors and shapes showing the sample IDs.}
	\label{fig: d5}
\end{figure*}

Spatial transcriptomics data from the human dorsolateral prefrontal cortex (DLPFC) are acquired using the 10x Visium platform. The dataset, accessible at \url{https://github.com/LieberInstitute/spatialLIBD}, comprises the information gathered from 12 postmortem DLPFC tissue sections obtained from three distinct neurotypical adult donors. 
Each sample's raw data encompasses 33538 genes.
The original study supplies manual annotations for the tissue layers, relying on cytoarchitecture. 
This facilitates the evaluation of spatial clustering by considering the manual annotations as the ground truth.
For each sample, we first remove spots with NAs in the annotations.
To make a fair comparison, we use the same data preprocessing step as DR.SC \citep{liu2022joint} and keep $q=15$.
For the proposed model TransST, we treat one sample as the target data and the remaining samples as the source data.
More specifically, we first preprocess each target data in the same way as described in DR.SC.
Then, we combine all the other samples by spots since they have the same number of genes.
Finally, the source data are subset by the genes in the target data.

Figures \ref{fig: d5}A and \ref{fig: d5}B depict the clustering results with and without the true number of clusters, respectively.
Our TransST maintains the highest ARI values in both settings and our spGMM ranks the second.
ARIs of GMM, SC-MEB, and DR.SC in Figure \ref{fig: d5}B are slightly lower due to the inaccurate estimation of the number of clusters. 
The clustering performance of our TransST is further demonstrated in the spatial heatmap in Figure \ref{fig: d5}C. 
Despite that none of the methods accurately estimate five clusters, our TransST method still achieves the highest ARI, and its spatial patterns match the true configuration most closely.
Spatial heatmaps for other samples are included in the Appendix.
The UMAP plots (Figure \ref{fig: d5}D) for TransST show that spots from different samples are well mixed, which implies the effectiveness of our data preprocessing procedure.

\subsection{TransST enables accurate identification of different areas in the mouse embryo}

\begin{figure*}[t!]
	\centering
	\includegraphics[width=0.98\linewidth,keepaspectratio]{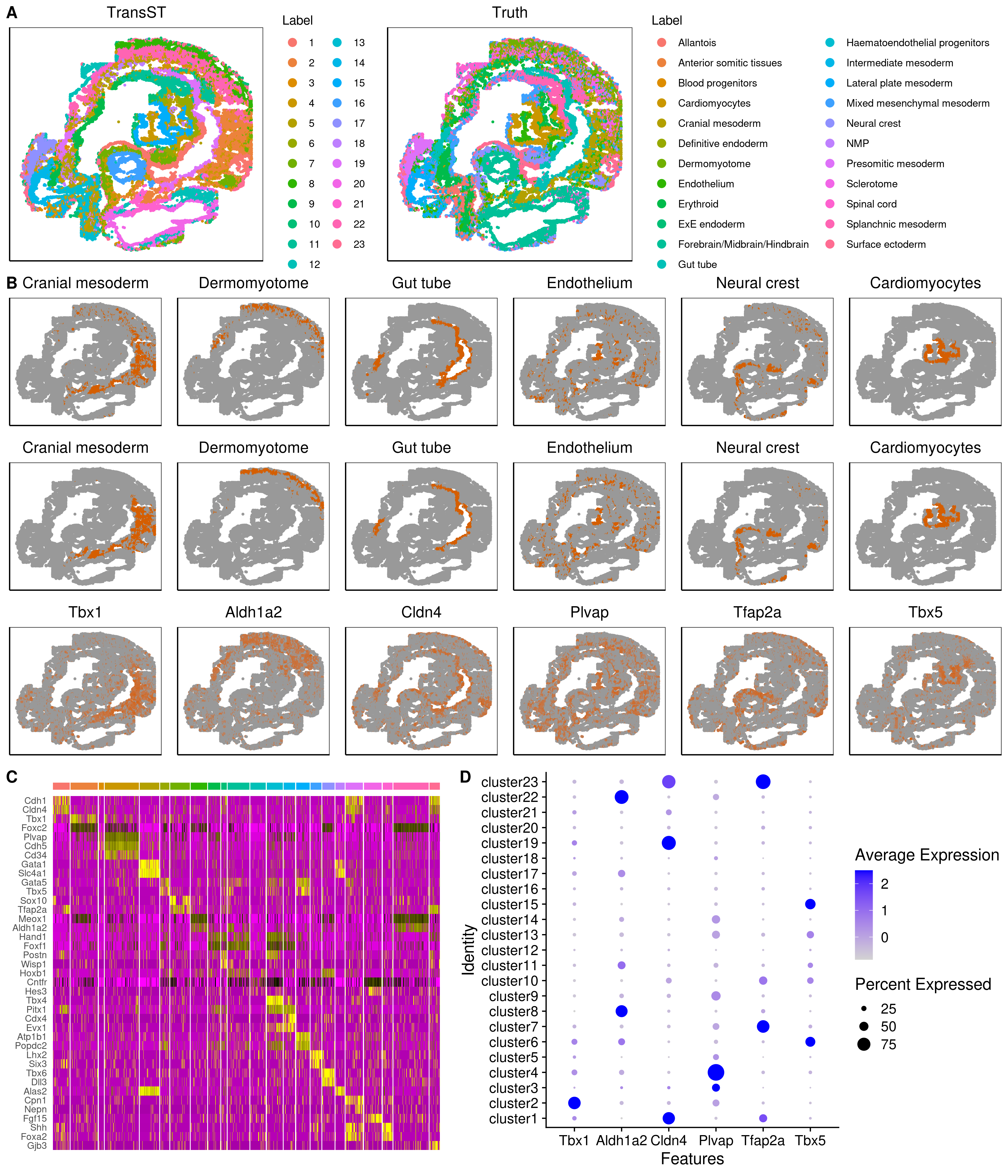}
	\caption{TransST enables accurate identification of different areas in the mouse embryo. 
		A: Spatial heatmaps colored by TransST and the original study (truth), respectively;
		B: Visualization of selected spatial domains identified by the original study (Row 1) and TransST (Row 2) and visualization of selected spatial domains by marker gene expressions identified by TransST (Row 3); 
		C: Heatmap of differentially expressed genes (top 2) identified by TransST; 
		D: Dot plot for the distributions of selected genes.}
	\label{fig: e1}
\end{figure*}

We acquire the high-resolution mouse embryos dataset from \url{https://content.cruk.cam.ac.uk/jmlab/SpatialMouseAtlas2020/}, generated through sequential fluorescence in situ hybridization (seqFISH). 
This dataset encompasses three samples with 19451, 14891, and 23194 spots, respectively.
All the samples share the same 351 genes.
After removing the low quality cells, 23 cell types are identified by referencing their closest neighbors in an existing scRNA-seq atlas (Gastrulation atlas).
We refer to \cite{lohoff2022integration} for more details about data preprocessing and cell type identification.
Using these manual annotations as a reference, we assess the clustering performances of TransST and spGMM in comparison to other (spatial) clustering methods. 
In the tandem analysis, we initially derive the top 15 principal components (PCs) from PCA and subsequently employ other spatial clustering methods based on these top 15 PCs.
To demonstrate the benefit of transfer learning, our TransST treats one sample as the target and the remaining samples as the source data.
While all the other methods only use the single target data in the analysis.

Figure \ref{fig: e1}A shows the spatial heatmap colored by clusters obtained from TransST and the original study (truth), respectively.
The spatial patterns by TransST match those from the original study very well, i.e., embryo areas identified by TransST are similar to the ground truth by the original paper.
To validate our findings, we select a few regions of the mouse embryo and check whether our estimation is consistent with the ground truth.
More specifically, we plot a few regions by the estimated clusters and highly expressed genes.
The first two rows of Figure \ref{fig: e1}B show the consistency in most regions. 
For example, the cardiomyocytes area identified by TransST (second row) is visually similar to the ground truth (first row).
Furthermore, the third row of Figure \ref{fig: e1}B shows that these regions can be identified by some differentially expressed genes as well.
For instance, we find that gene Tbx1 uniquely determines cranial mesoderm (cluster 2) in the original study, as shown on the dot plot of Figure \ref{fig: e1}D, and Tbx1 dominates the percent expressed.
Tbx1 serves as a crucial regulator of esophagus striated muscles, which are identified as a third derivative of cardiopharyngeal mesoderm contributing to both second heart field derivatives and head muscles \citep{gopalakrishnan2015cranial}.
We also found that gene Aldh1a2 is highly expressed in the area of dermomyotome, which is identified as clusters 8 and 22 using the proposed TransST (see Figure \ref{fig: e1}D).
Moreover, we provide more details on the DEGs in Figure \ref{fig: e1}C and Figure \ref{fig: e1}D.
There is a visually clear pattern in the heatmap. 
Most identified genes are highly expressed in a single cluster or a few clusters.
For example, genes Gata1, Slc4a1, Alas2 are highly expressed in Cluster 5 while other genes exhibit low expression levels (almost 0).
Most of these identified genes have also been well studied in the literature. 
For instance, the role of Cdh1 in the mouse uterus is indispensable for endometrial differentiation, gland development, and the adult function of the uterus \citep{reardon2012cdh1}.
Mice show a drastic bloodless phenotype upon disruption of the GATA1 gene \citep{kobayashi2007regulation}.
Gene Tbx4 is expressed very strongly in leg \citep{isaac1998tbx} and gene Lhx2 plays a crucial role in the normal development of the eye, cerebral cortex, and the efficient progression of definitive erythropoiesis \citep{porter1997lhx2}.
The dot plot depicts the contributions of the selected genes for each cluster.
For example, gene Plvap is highly expressed in Cluster 4 while gene Tfap2a may highly correlate with Clusters 7 and 23.

\subsection{TransST enables identification of different regions and differentially expressed genes in the squamous cell carcinoma data}

\begin{figure*}[hbt!]
	\centering
	\includegraphics[width=\linewidth,keepaspectratio]{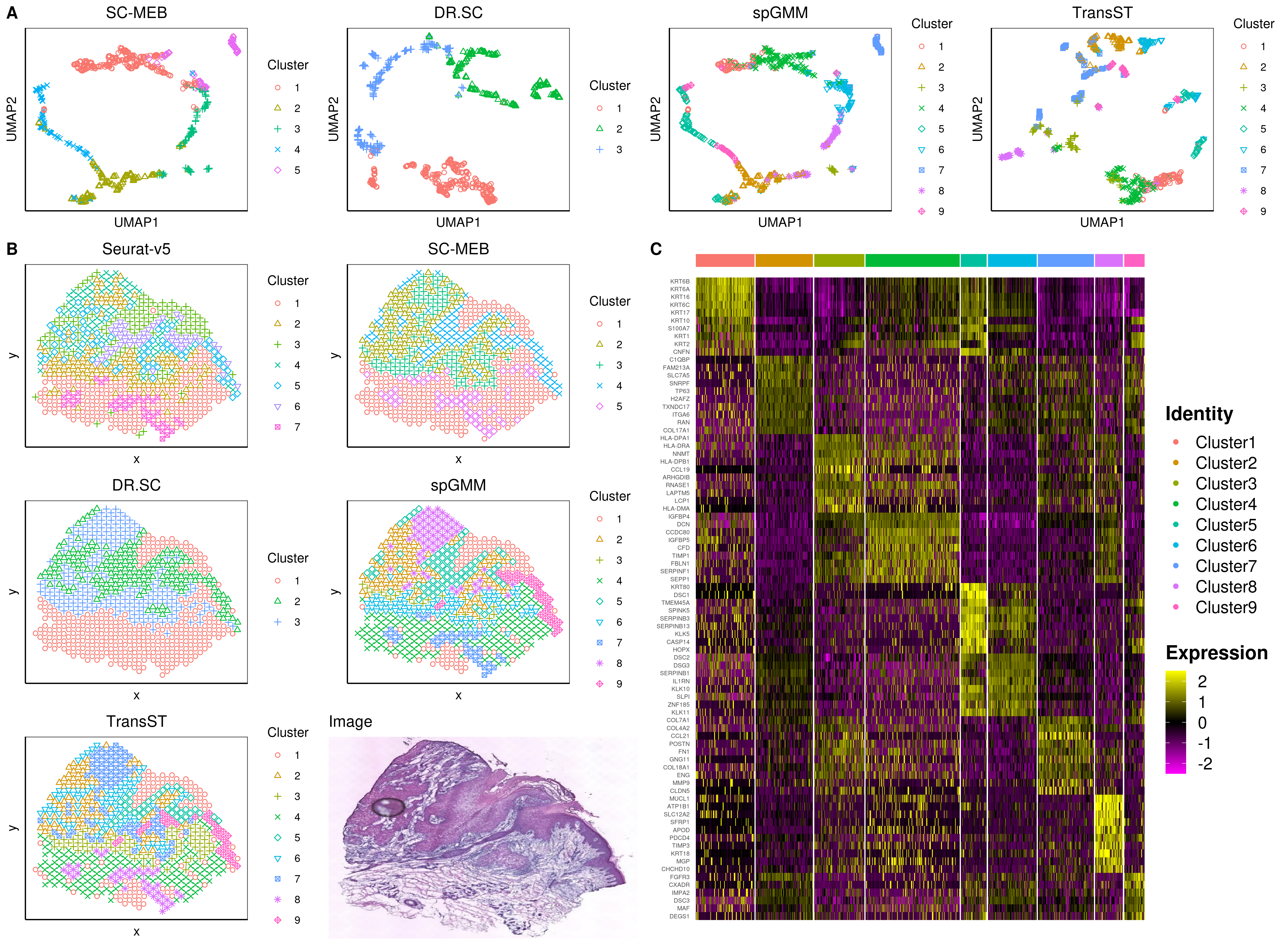}
	\caption{TransST enables identification of different regions and differentially expressed genes in the squamous cell carcinoma data. 
		A: UMAP for the target sample by the spatial methods SC-MEB, DR.SC, spGMM, and TransST;
		B: Spatial heatmaps by various clustering methods and the scanned image;
		C: Heatmap of top 10 differentially expressed genes for each cell type identified by TransST.
	}
	\label{fig: s1}
\end{figure*}

Epithelial cancers constitute approximately 90\% of human malignancies. 
Cutaneous squamous cell carcinoma (cSCC) stands out as a prototype within this category, showcasing distinctive features like the disruption of tissue polarity and invasion of the basement membrane \citep{ji2020multimodal}.
Single-cell transcriptomics encounters challenges due to the loss of spatial information during tissue dissociation but has unveiled intra-tumoral heterogeneity in numerous cancer types.
By transferring information from the scRNA-seq data, we aim to facilitate the dissection of cellular communication of the tumor microenvironment.

The squamous cell carcinoma dataset is publicly available from \url{https://github.com/drieslab/spatial-datasets/tree/master/data/2020_ST_SCC}, which is abbreviated as ST-SCC in our analysis.
It contains one scRNA-seq sample and one spatial transcriptomics sample.
The scRNA-seq sample has 6734 cells and 7287 genes with the annotated labels.
In our analysis, we treat this scRNA-seq sample as the source data since it contains annotated cell types.
The spatial transcriptomics sample (slide 1) has 666 spots and 17138 genes.
It is treated as the target data and cell types are not publicly available.
For our TransST, genes in the source data are chosen directly from the target data.
All the methods use the same $q=15$.

Figure \ref{fig: s1}B depicts the spatial heatmap by different methods.
Comparing our TransST and spGMM to the spatial methods SC-MEB, DR.SC, and Seurat, our method chooses a larger $K$ and the spatial heatmaps match better to the raw image.
\cite{ji2020multimodal} identifies 7 and 11 clusters in similar samples. 
UMAPs of Figure \ref{fig: s1}A also show that TransST separates cells better in the low-dimensional space. 
Figure \ref{fig: s1}C shows good separation of the differentially expressed genes across different cell types.
We next study some identified genes based on the existing biomedical literature.
For example, clusters with the highly expressed epithelial-associated genes such as KRT5, KRT14, KRT1, and KRT10 are inferred to be epithelial cells for Cluster 1.
MHC class II genes (e.g., HLA-DPA1, HLA-DRA, and HLA-DPB1) are found to be important for myeloid cells.
Therefore, Cluster 3 is inferred as myeloid cells.
Genes COL7A1 and COL4A2 are inferred to fibroblasts since they are highly expressed in Cluster 7.
The abovementioned analyses are consistent with findings in the original study \citep{ji2020multimodal}.
Beyond these well-studied genes in the literature, we also have some new findings by our method.
For instance,
Cluster 2 highly expresses genes C1QBP, FAM213A, and SLC7A5, which are rarely investigated in the literature.
Gene FAM213A is found to protect cells from oxidative stress and modulate osteoclast differentiation \citep{xu2010pamm}.
Gene SLC7A5 is upregulated in actinic keratosis and squamous cell carcinoma \citep{o2019targeting}.
The synergistic effects warrant further investigation.
Besides, DSC1 is significantly associated with infiltration by most immune cells \citep{luan2021identification};
HOPX expression is linked to immune processes and shows high enrichment of T cells, suggesting it could serve as an immune checkpoint in skin cutaneous melanoma \citep{he2023hopx}.
Thus, we infer Cluster 5 as immune cells.
The expression of SFRP1 is notably lower in cSCC tissues compared to tissues from control subjects, with statistical significance (p-value $<$ 0.05) \citep{halifu2016wnt1};
Gene APOD may play continuous roles in cSCC development \citep{chen2021seven}.
Based on these observations, we infer Cluster 8 as prognostic cells.
Other genes such as KLK10 and FGFR3 are also highly expressed in some clusters, but their functionalities in cSCC are rarely investigated in the literature.
Our findings may provide some insights for biologists and geneticists to further explore cSCC data.

\section{Discussion}
\label{sec: dis}

Most existing methods for spatial transcriptomics data simply focus on the target data, while ours has the capability of utilizing external information from other similar or related studies.
For example, scRNA-seq data, with its swiftly expanding repositories of annotated public data and RNA-focused computational tools capable of delivering precise classifications, serves as a natural source domain for transferring information to other modalities.
Motivated by the rich information encoded in the source data, we leverage the power of scRNA-seq data as a source and spatial transcriptomics data as a target to develop a transfer learning method that surpasses existing approaches. 
Our method incorporates external information, providing a more comprehensive and accurate representation of cellular dynamics within spatial contexts. 
In this discussion, we delve into the key aspects of our methodology, compare it with existing methods, and explore the implications and potential advancements in the field.

The integration of scRNA-seq and spatial transcriptomics data is critical for capturing both the cellular heterogeneity at the single-cell level and the spatial organization of cells within tissues. 
By combining these two types of data, our method exploits the complementary strengths of the technologies. scRNA-seq contains rich gene expression information for the identification of distinct cell types and states, while spatial transcriptomics provides additional valuable spatial information, enabling us to infer the spatial distribution of these cell types.

Our comprehensive evaluation and comparisons with existing methods demonstrate the superior performance of our method. 
Our method is shown to have robust performance in various types of data, including human breast tumor data, human brain data, mouse embryo data, and squamous cell carcinoma data.
We outperform the competitors in accurately reconstructing spatial spot distributions, identifying cell subpopulations, and capturing subtle spatial patterns. 
The incorporation of external information plays a pivotal role in achieving these advancements, emphasizing the importance of transferring external knowledge from relevant other studies to boost the discovery and interpretation of spatial transcriptomics data.

In summary, we develop a novel transfer learning method for delineating cell hetergeneity in spatial transcriptomics data and show the competing performance in comparison with the existing popular methods. 
While our method represents a significant step forward to improve learning in spatial transcriptomics data by transferring useful information from externally relevant studies, there are still opportunities for refinement and expansion.
For example, the source and target data may not align well, often exhibiting significantly different distributions. 
Developing an algorithm capable of effectively aligning the source and target data can enhance the performance of transfer learning in spatial transcriptomics data analysis.

\backmatter

\bmhead{Supplementary information}

We include details on the selection of number of clusters $K$, simulation settings, real data preprocessing, mathematical derivation of TransST, and more experiment results in the supplementary file.




\section*{Declarations}


\begin{itemize}
\item Ethics approval and consent to participate: Not applicable.
\item Consent for publication: Not applicable.
\item Availability of data and materials: 
HER2 positive breast tumor dataset is obtained from \url{https://github.com/almaan/her2st}.
DLPFC dataset is accessible at \url{https://github.com/LieberInstitute/spatialLIBD}.
Mouse embryo dataset is acquired from \url{https://content.cruk.cam.ac.uk/jmlab/SpatialMouseAtlas2020/}.
The squamous cell carcinoma dataset is publicly available from \url{https://github.com/drieslab/spatial-datasets/tree/master/data/2020_ST_SCC}.The source code is publicly available at \url{https://github.com/shuoshuoliu/TransST}.
\item Competing interests: The authors declare that they have no competing interests.
\item Funding: The work was partially supported by the National Institute of Health Grants NCI 5P30 CA013696, NCI P01 CA098101, and NIAID 1R01 AI143886.
\item Authors' contributions: S.S.L., S.W., M.Y., and J.H. conceptualized the study. 
S.S.L was responsible for algorithm development and implementation. 
S.S.L. initially wrote the manuscript and analyzed the multiple datasets, S.W. helped the data analysis, and J.H. supervised the whole project.
S.S.L., S.W. and J.H. jointly analyzed the data. 
Y.C. helped running codes during revision.
All authors contributed to the writing of the paper.
\item Acknowledgements: Not applicable.
\end{itemize}



\bibliography{sn-bibliography}


\begin{thebibliography}{43}
\ifx \bisbn   \undefined \def \bisbn  #1{ISBN #1}\fi
\ifx \binits  \undefined \def \binits#1{#1}\fi
\ifx \bauthor  \undefined \def \bauthor#1{#1}\fi
\ifx \batitle  \undefined \def \batitle#1{#1}\fi
\ifx \bjtitle  \undefined \def \bjtitle#1{#1}\fi
\ifx \bvolume  \undefined \def \bvolume#1{\textbf{#1}}\fi
\ifx \byear  \undefined \def \byear#1{#1}\fi
\ifx \bissue  \undefined \def \bissue#1{#1}\fi
\ifx \bfpage  \undefined \def \bfpage#1{#1}\fi
\ifx \blpage  \undefined \def \blpage #1{#1}\fi
\ifx \burl  \undefined \def \burl#1{\textsf{#1}}\fi
\ifx \doiurl  \undefined \def \doiurl#1{\url{https://doi.org/#1}}\fi
\ifx \betal  \undefined \def \betal{\textit{et al.}}\fi
\ifx \binstitute  \undefined \def \binstitute#1{#1}\fi
\ifx \binstitutionaled  \undefined \def \binstitutionaled#1{#1}\fi
\ifx \bctitle  \undefined \def \bctitle#1{#1}\fi
\ifx \beditor  \undefined \def \beditor#1{#1}\fi
\ifx \bpublisher  \undefined \def \bpublisher#1{#1}\fi
\ifx \bbtitle  \undefined \def \bbtitle#1{#1}\fi
\ifx \bedition  \undefined \def \bedition#1{#1}\fi
\ifx \bseriesno  \undefined \def \bseriesno#1{#1}\fi
\ifx \blocation  \undefined \def \blocation#1{#1}\fi
\ifx \bsertitle  \undefined \def \bsertitle#1{#1}\fi
\ifx \bsnm \undefined \def \bsnm#1{#1}\fi
\ifx \bsuffix \undefined \def \bsuffix#1{#1}\fi
\ifx \bparticle \undefined \def \bparticle#1{#1}\fi
\ifx \barticle \undefined \def \barticle#1{#1}\fi
\bibcommenthead
\ifx \bconfdate \undefined \def \bconfdate #1{#1}\fi
\ifx \botherref \undefined \def \botherref #1{#1}\fi
\ifx \url \undefined \def \url#1{\textsf{#1}}\fi
\ifx \bchapter \undefined \def \bchapter#1{#1}\fi
\ifx \bbook \undefined \def \bbook#1{#1}\fi
\ifx \bcomment \undefined \def \bcomment#1{#1}\fi
\ifx \oauthor \undefined \def \oauthor#1{#1}\fi
\ifx \citeauthoryear \undefined \def \citeauthoryear#1{#1}\fi
\ifx \endbibitem  \undefined \def \endbibitem {}\fi
\ifx \bconflocation  \undefined \def \bconflocation#1{#1}\fi
\ifx \arxivurl  \undefined \def \arxivurl#1{\textsf{#1}}\fi
\csname PreBibitemsHook\endcsname

\bibitem[\protect\citeauthoryear{Yang et~al.}{2022}]{yang2022sc}
\begin{barticle}
\bauthor{\bsnm{Yang}, \binits{Y.}},
\bauthor{\bsnm{Shi}, \binits{X.}},
\bauthor{\bsnm{Liu}, \binits{W.}},
\bauthor{\bsnm{Zhou}, \binits{Q.}},
\bauthor{\bsnm{Chan~Lau}, \binits{M.}},
\bauthor{\bsnm{Chun Tatt~Lim}, \binits{J.}},
\bauthor{\bsnm{Sun}, \binits{L.}},
\bauthor{\bsnm{Ng}, \binits{C.C.Y.}},
\bauthor{\bsnm{Yeong}, \binits{J.}},
\bauthor{\bsnm{Liu}, \binits{J.}}:
\batitle{Sc-meb: spatial clustering with hidden markov random field using empirical bayes}.
\bjtitle{Briefings in Bioinformatics}
\bvolume{23}(\bissue{1}),
\bfpage{466}
(\byear{2022})
\end{barticle}
\endbibitem

\bibitem[\protect\citeauthoryear{Lin et~al.}{2022}]{lin2022scjoint}
\begin{barticle}
\bauthor{\bsnm{Lin}, \binits{Y.}},
\bauthor{\bsnm{Wu}, \binits{T.-Y.}},
\bauthor{\bsnm{Wan}, \binits{S.}},
\bauthor{\bsnm{Yang}, \binits{J.Y.}},
\bauthor{\bsnm{Wong}, \binits{W.H.}},
\bauthor{\bsnm{Wang}, \binits{Y.R.}}:
\batitle{scjoint integrates atlas-scale single-cell rna-seq and atac-seq data with transfer learning}.
\bjtitle{Nature Biotechnology}
\bvolume{40}(\bissue{5}),
\bfpage{703}--\blpage{710}
(\byear{2022})
\end{barticle}
\endbibitem

\bibitem[\protect\citeauthoryear{Lohoff et~al.}{2022}]{lohoff2022integration}
\begin{barticle}
\bauthor{\bsnm{Lohoff}, \binits{T.}},
\bauthor{\bsnm{Ghazanfar}, \binits{S.}},
\bauthor{\bsnm{Missarova}, \binits{A.}},
\bauthor{\bsnm{Koulena}, \binits{N.}},
\bauthor{\bsnm{Pierson}, \binits{N.}},
\bauthor{\bsnm{Griffiths}, \binits{J.}},
\bauthor{\bsnm{Bardot}, \binits{E.}},
\bauthor{\bsnm{Eng}, \binits{C.-H.}},
\bauthor{\bsnm{Tyser}, \binits{R.}},
\bauthor{\bsnm{Argelaguet}, \binits{R.}}, \betal:
\batitle{Integration of spatial and single-cell transcriptomic data elucidates mouse organogenesis}.
\bjtitle{Nature Biotechnology}
\bvolume{40}(\bissue{1}),
\bfpage{74}--\blpage{85}
(\byear{2022})
\end{barticle}
\endbibitem

\bibitem[\protect\citeauthoryear{Ji et~al.}{2020}]{ji2020multimodal}
\begin{barticle}
\bauthor{\bsnm{Ji}, \binits{A.L.}},
\bauthor{\bsnm{Rubin}, \binits{A.J.}},
\bauthor{\bsnm{Thrane}, \binits{K.}},
\bauthor{\bsnm{Jiang}, \binits{S.}},
\bauthor{\bsnm{Reynolds}, \binits{D.L.}},
\bauthor{\bsnm{Meyers}, \binits{R.M.}},
\bauthor{\bsnm{Guo}, \binits{M.G.}},
\bauthor{\bsnm{George}, \binits{B.M.}},
\bauthor{\bsnm{Mollbrink}, \binits{A.}},
\bauthor{\bsnm{Bergenstr{\aa}hle}, \binits{J.}}, \betal:
\batitle{Multimodal analysis of composition and spatial architecture in human squamous cell carcinoma}.
\bjtitle{Cell}
\bvolume{182}(\bissue{2}),
\bfpage{497}--\blpage{514}
(\byear{2020})
\end{barticle}
\endbibitem

\bibitem[\protect\citeauthoryear{Hu et~al.}{2021}]{hu2021spagcn}
\begin{barticle}
\bauthor{\bsnm{Hu}, \binits{J.}},
\bauthor{\bsnm{Li}, \binits{X.}},
\bauthor{\bsnm{Coleman}, \binits{K.}},
\bauthor{\bsnm{Schroeder}, \binits{A.}},
\bauthor{\bsnm{Ma}, \binits{N.}},
\bauthor{\bsnm{Irwin}, \binits{D.J.}},
\bauthor{\bsnm{Lee}, \binits{E.B.}},
\bauthor{\bsnm{Shinohara}, \binits{R.T.}},
\bauthor{\bsnm{Li}, \binits{M.}}:
\batitle{Spagcn: Integrating gene expression, spatial location and histology to identify spatial domains and spatially variable genes by graph convolutional network}.
\bjtitle{Nature Methods}
\bvolume{18}(\bissue{11}),
\bfpage{1342}--\blpage{1351}
(\byear{2021})
\end{barticle}
\endbibitem

\bibitem[\protect\citeauthoryear{Dries et~al.}{2021}]{dries2021giotto}
\begin{barticle}
\bauthor{\bsnm{Dries}, \binits{R.}},
\bauthor{\bsnm{Zhu}, \binits{Q.}},
\bauthor{\bsnm{Dong}, \binits{R.}},
\bauthor{\bsnm{Eng}, \binits{C.-H.L.}},
\bauthor{\bsnm{Li}, \binits{H.}},
\bauthor{\bsnm{Liu}, \binits{K.}},
\bauthor{\bsnm{Fu}, \binits{Y.}},
\bauthor{\bsnm{Zhao}, \binits{T.}},
\bauthor{\bsnm{Sarkar}, \binits{A.}},
\bauthor{\bsnm{Bao}, \binits{F.}}, \betal:
\batitle{Giotto: a toolbox for integrative analysis and visualization of spatial expression data}.
\bjtitle{Genome Biology}
\bvolume{22},
\bfpage{1}--\blpage{31}
(\byear{2021})
\end{barticle}
\endbibitem

\bibitem[\protect\citeauthoryear{Zhao et~al.}{2021}]{zhao2021spatial}
\begin{barticle}
\bauthor{\bsnm{Zhao}, \binits{E.}},
\bauthor{\bsnm{Stone}, \binits{M.R.}},
\bauthor{\bsnm{Ren}, \binits{X.}},
\bauthor{\bsnm{Guenthoer}, \binits{J.}},
\bauthor{\bsnm{Smythe}, \binits{K.S.}},
\bauthor{\bsnm{Pulliam}, \binits{T.}},
\bauthor{\bsnm{Williams}, \binits{S.R.}},
\bauthor{\bsnm{Uytingco}, \binits{C.R.}},
\bauthor{\bsnm{Taylor}, \binits{S.E.}},
\bauthor{\bsnm{Nghiem}, \binits{P.}}, \betal:
\batitle{Spatial transcriptomics at subspot resolution with bayesspace}.
\bjtitle{Nature Biotechnology}
\bvolume{39}(\bissue{11}),
\bfpage{1375}--\blpage{1384}
(\byear{2021})
\end{barticle}
\endbibitem

\bibitem[\protect\citeauthoryear{Markos et~al.}{2019}]{markos2019beyond}
\begin{barticle}
\bauthor{\bsnm{Markos}, \binits{A.}},
\bauthor{\bsnm{D'Enza}, \binits{A.I.}},
\bauthor{\bsnm{Velden}, \binits{M.}}:
\batitle{Beyond tandem analysis: Joint dimension reduction and clustering in r}.
\bjtitle{Journal of Statistical Software}
\bvolume{91},
\bfpage{1}--\blpage{24}
(\byear{2019})
\end{barticle}
\endbibitem

\bibitem[\protect\citeauthoryear{Aggarwal and Yu}{2000}]{aggarwal2000finding}
\begin{bchapter}
\bauthor{\bsnm{Aggarwal}, \binits{C.C.}},
\bauthor{\bsnm{Yu}, \binits{P.S.}}:
\bctitle{Finding generalized projected clusters in high dimensional spaces}.
In: \bbtitle{Proceedings of the 2000 ACM SIGMOD International Conference on Management of Data},
pp. \bfpage{70}--\blpage{81}
(\byear{2000})
\end{bchapter}
\endbibitem

\bibitem[\protect\citeauthoryear{Liu et~al.}{2022}]{liu2022joint}
\begin{barticle}
\bauthor{\bsnm{Liu}, \binits{W.}},
\bauthor{\bsnm{Liao}, \binits{X.}},
\bauthor{\bsnm{Yang}, \binits{Y.}},
\bauthor{\bsnm{Lin}, \binits{H.}},
\bauthor{\bsnm{Yeong}, \binits{J.}},
\bauthor{\bsnm{Zhou}, \binits{X.}},
\bauthor{\bsnm{Shi}, \binits{X.}},
\bauthor{\bsnm{Liu}, \binits{J.}}:
\batitle{Joint dimension reduction and clustering analysis of single-cell rna-seq and spatial transcriptomics data}.
\bjtitle{Nucleic Acids Research}
\bvolume{50}(\bissue{12}),
\bfpage{72}--\blpage{72}
(\byear{2022})
\end{barticle}
\endbibitem

\bibitem[\protect\citeauthoryear{Mieth et~al.}{2019}]{mieth2019using}
\begin{barticle}
\bauthor{\bsnm{Mieth}, \binits{B.}},
\bauthor{\bsnm{Hockley}, \binits{J.R.}},
\bauthor{\bsnm{G{\"o}rnitz}, \binits{N.}},
\bauthor{\bsnm{Vidovic}, \binits{M.M.-C.}},
\bauthor{\bsnm{M{\"u}ller}, \binits{K.-R.}},
\bauthor{\bsnm{Gutteridge}, \binits{A.}},
\bauthor{\bsnm{Ziemek}, \binits{D.}}:
\batitle{Using transfer learning from prior reference knowledge to improve the clustering of single-cell rna-seq data}.
\bjtitle{Scientific Reports}
\bvolume{9}(\bissue{1}),
\bfpage{20353}
(\byear{2019})
\end{barticle}
\endbibitem

\bibitem[\protect\citeauthoryear{Peng et~al.}{2021}]{peng2021integration}
\begin{barticle}
\bauthor{\bsnm{Peng}, \binits{M.}},
\bauthor{\bsnm{Li}, \binits{Y.}},
\bauthor{\bsnm{Wamsley}, \binits{B.}},
\bauthor{\bsnm{Wei}, \binits{Y.}},
\bauthor{\bsnm{Roeder}, \binits{K.}}:
\batitle{Integration and transfer learning of single-cell transcriptomes via cfit}.
\bjtitle{Proceedings of the National Academy of Sciences}
\bvolume{118}(\bissue{10}),
\bfpage{2024383118}
(\byear{2021})
\end{barticle}
\endbibitem

\bibitem[\protect\citeauthoryear{Hu et~al.}{2020}]{hu2020iterative}
\begin{barticle}
\bauthor{\bsnm{Hu}, \binits{J.}},
\bauthor{\bsnm{Li}, \binits{X.}},
\bauthor{\bsnm{Hu}, \binits{G.}},
\bauthor{\bsnm{Lyu}, \binits{Y.}},
\bauthor{\bsnm{Susztak}, \binits{K.}},
\bauthor{\bsnm{Li}, \binits{M.}}:
\batitle{Iterative transfer learning with neural network for clustering and cell type classification in single-cell rna-seq analysis}.
\bjtitle{Nature Machine Intelligence}
\bvolume{2}(\bissue{10}),
\bfpage{607}--\blpage{618}
(\byear{2020})
\end{barticle}
\endbibitem

\bibitem[\protect\citeauthoryear{Hao et~al.}{2023}]{hao2023dictionary}
\begin{botherref}
\oauthor{\bsnm{Hao}, \binits{Y.}},
\oauthor{\bsnm{Stuart}, \binits{T.}},
\oauthor{\bsnm{Kowalski}, \binits{M.H.}},
\oauthor{\bsnm{Choudhary}, \binits{S.}},
\oauthor{\bsnm{Hoffman}, \binits{P.}},
\oauthor{\bsnm{Hartman}, \binits{A.}},
\oauthor{\bsnm{Srivastava}, \binits{A.}},
\oauthor{\bsnm{Molla}, \binits{G.}},
\oauthor{\bsnm{Madad}, \binits{S.}},
\oauthor{\bsnm{Fernandez-Granda}, \binits{C.}}, et al.:
Dictionary learning for integrative, multimodal and scalable single-cell analysis.
Nature Biotechnology,
1--12
(2023)
\end{botherref}
\endbibitem

\bibitem[\protect\citeauthoryear{Blondel et~al.}{2008}]{blondel2008fast}
\begin{barticle}
\bauthor{\bsnm{Blondel}, \binits{V.D.}},
\bauthor{\bsnm{Guillaume}, \binits{J.-L.}},
\bauthor{\bsnm{Lambiotte}, \binits{R.}},
\bauthor{\bsnm{Lefebvre}, \binits{E.}}:
\batitle{Fast unfolding of communities in large networks}.
\bjtitle{Journal of Statistical Mechanics: Theory and Experiment}
\bvolume{2008}(\bissue{10}),
\bfpage{10008}
(\byear{2008})
\end{barticle}
\endbibitem

\bibitem[\protect\citeauthoryear{Traag et~al.}{2019}]{traag2019louvain}
\begin{barticle}
\bauthor{\bsnm{Traag}, \binits{V.A.}},
\bauthor{\bsnm{Waltman}, \binits{L.}},
\bauthor{\bsnm{Van~Eck}, \binits{N.J.}}:
\batitle{From louvain to leiden: guaranteeing well-connected communities}.
\bjtitle{Scientific reports}
\bvolume{9}(\bissue{1}),
\bfpage{5233}
(\byear{2019})
\end{barticle}
\endbibitem

\bibitem[\protect\citeauthoryear{Long et~al.}{2023}]{long2023spatially}
\begin{barticle}
\bauthor{\bsnm{Long}, \binits{Y.}},
\bauthor{\bsnm{Ang}, \binits{K.S.}},
\bauthor{\bsnm{Li}, \binits{M.}},
\bauthor{\bsnm{Chong}, \binits{K.L.K.}},
\bauthor{\bsnm{Sethi}, \binits{R.}},
\bauthor{\bsnm{Zhong}, \binits{C.}},
\bauthor{\bsnm{Xu}, \binits{H.}},
\bauthor{\bsnm{Ong}, \binits{Z.}},
\bauthor{\bsnm{Sachaphibulkij}, \binits{K.}},
\bauthor{\bsnm{Chen}, \binits{A.}}, \betal:
\batitle{Spatially informed clustering, integration, and deconvolution of spatial transcriptomics with graphst}.
\bjtitle{Nature Communications}
\bvolume{14}(\bissue{1}),
\bfpage{1155}
(\byear{2023})
\end{barticle}
\endbibitem

\bibitem[\protect\citeauthoryear{Andersson et~al.}{2021}]{andersson2021spatial}
\begin{barticle}
\bauthor{\bsnm{Andersson}, \binits{A.}},
\bauthor{\bsnm{Larsson}, \binits{L.}},
\bauthor{\bsnm{Stenbeck}, \binits{L.}},
\bauthor{\bsnm{Salm{\'e}n}, \binits{F.}},
\bauthor{\bsnm{Ehinger}, \binits{A.}},
\bauthor{\bsnm{Wu}, \binits{S.Z.}},
\bauthor{\bsnm{Al-Eryani}, \binits{G.}},
\bauthor{\bsnm{Roden}, \binits{D.}},
\bauthor{\bsnm{Swarbrick}, \binits{A.}},
\bauthor{\bsnm{Borg}, \binits{{\AA}.}}, \betal:
\batitle{Spatial deconvolution of her2-positive breast cancer delineates tumor-associated cell type interactions}.
\bjtitle{Nature Communications}
\bvolume{12}(\bissue{1}),
\bfpage{6012}
(\byear{2021})
\end{barticle}
\endbibitem

\bibitem[\protect\citeauthoryear{Wu et~al.}{2020}]{wu2020comprehensive}
\begin{barticle}
\bauthor{\bsnm{Wu}, \binits{J.}},
\bauthor{\bsnm{Luo}, \binits{M.}},
\bauthor{\bsnm{Chen}, \binits{Z.}},
\bauthor{\bsnm{Huang}, \binits{X.}}:
\batitle{Comprehensive analysis of cd2 in the immune microenvironment of breast cancer}.
\bjtitle{Revista Argentina de Cl{\'\i}nica Psicol{\'o}gica}
\bvolume{29}(\bissue{3}),
\bfpage{1249}--\blpage{1256}
(\byear{2020})
\end{barticle}
\endbibitem

\bibitem[\protect\citeauthoryear{Chen et~al.}{2021}]{chen2021cd2}
\begin{barticle}
\bauthor{\bsnm{Chen}, \binits{Y.}},
\bauthor{\bsnm{Meng}, \binits{Z.}},
\bauthor{\bsnm{Zhang}, \binits{L.}},
\bauthor{\bsnm{Liu}, \binits{F.}}:
\batitle{Cd2 is a novel immune-related prognostic biomarker of invasive breast carcinoma that modulates the tumor microenvironment}.
\bjtitle{Frontiers in Immunology}
\bvolume{12},
\bfpage{664845}
(\byear{2021})
\end{barticle}
\endbibitem

\bibitem[\protect\citeauthoryear{Vizoso et~al.}{2001}]{vizoso2001lysozyme}
\begin{barticle}
\bauthor{\bsnm{Vizoso}, \binits{F.}},
\bauthor{\bsnm{Plaza}, \binits{E.}},
\bauthor{\bsnm{V{\'a}zquez}, \binits{J.}},
\bauthor{\bsnm{Serra}, \binits{C.}},
\bauthor{\bsnm{Lamelas}, \binits{M.L.}},
\bauthor{\bsnm{Gonz{\'a}lez}, \binits{L.O.}},
\bauthor{\bsnm{Merino}, \binits{A.M.}},
\bauthor{\bsnm{M{\'e}ndez}, \binits{J.}}:
\batitle{Lysozyme expression by breast carcinomas, correlation with clinicopathologic parameters, and prognostic significance}.
\bjtitle{Annals of surgical oncology}
\bvolume{8},
\bfpage{667}--\blpage{674}
(\byear{2001})
\end{barticle}
\endbibitem

\bibitem[\protect\citeauthoryear{Serra et~al.}{2002}]{serra2002expression}
\begin{barticle}
\bauthor{\bsnm{Serra}, \binits{C.}},
\bauthor{\bsnm{Vizoso}, \binits{F.}},
\bauthor{\bsnm{Alonso}, \binits{L.}},
\bauthor{\bsnm{Rodr{\'\i}guez}, \binits{J.C.}},
\bauthor{\bsnm{Gonz{\'a}lez}, \binits{L.O.}},
\bauthor{\bsnm{Fern{\'a}ndez}, \binits{M.}},
\bauthor{\bsnm{Lamelas}, \binits{M.L.}},
\bauthor{\bsnm{S{\'a}nchez}, \binits{L.M.}},
\bauthor{\bsnm{Garc{\'\i}a-Mu{\~n}iz}, \binits{J.L.}},
\bauthor{\bsnm{Baltasar}, \binits{A.}}, \betal:
\batitle{Expression and prognostic significance of lysozyme in male breast cancer}.
\bjtitle{Breast Cancer Research}
\bvolume{4},
\bfpage{1}--\blpage{8}
(\byear{2002})
\end{barticle}
\endbibitem

\bibitem[\protect\citeauthoryear{Tietscher et~al.}{2023}]{tietscher2023comprehensive}
\begin{barticle}
\bauthor{\bsnm{Tietscher}, \binits{S.}},
\bauthor{\bsnm{Wagner}, \binits{J.}},
\bauthor{\bsnm{Anzeneder}, \binits{T.}},
\bauthor{\bsnm{Langwieder}, \binits{C.}},
\bauthor{\bsnm{Rees}, \binits{M.}},
\bauthor{\bsnm{Sobottka}, \binits{B.}},
\bauthor{\bsnm{Souza}, \binits{N.}},
\bauthor{\bsnm{Bodenmiller}, \binits{B.}}:
\batitle{A comprehensive single-cell map of t cell exhaustion-associated immune environments in human breast cancer}.
\bjtitle{Nature Communications}
\bvolume{14}(\bissue{1}),
\bfpage{98}
(\byear{2023})
\end{barticle}
\endbibitem

\bibitem[\protect\citeauthoryear{Woo et~al.}{2019}]{woo2019promoter}
\begin{barticle}
\bauthor{\bsnm{Woo}, \binits{H.Y.}},
\bauthor{\bsnm{Do}, \binits{S.-I.}},
\bauthor{\bsnm{Kim}, \binits{S.H.}},
\bauthor{\bsnm{Song}, \binits{S.Y.}},
\bauthor{\bsnm{Kim}, \binits{H.-S.}}:
\batitle{Promoter methylation down-regulates b-cell translocation gene 1 expression in breast carcinoma}.
\bjtitle{Anticancer Research}
\bvolume{39}(\bissue{10}),
\bfpage{5361}--\blpage{5367}
(\byear{2019})
\end{barticle}
\endbibitem

\bibitem[\protect\citeauthoryear{Amirfallah et~al.}{2019}]{amirfallah2019high}
\begin{barticle}
\bauthor{\bsnm{Amirfallah}, \binits{A.}},
\bauthor{\bsnm{Arason}, \binits{A.}},
\bauthor{\bsnm{Einarsson}, \binits{H.}},
\bauthor{\bsnm{Gudmundsdottir}, \binits{E.T.}},
\bauthor{\bsnm{Freysteinsdottir}, \binits{E.S.}},
\bauthor{\bsnm{Olafsdottir}, \binits{K.A.}},
\bauthor{\bsnm{Johannsson}, \binits{O.T.}},
\bauthor{\bsnm{Agnarsson}, \binits{B.A.}},
\bauthor{\bsnm{Barkardottir}, \binits{R.B.}},
\bauthor{\bsnm{Reynisdottir}, \binits{I.}}:
\batitle{High expression of the vacuole membrane protein 1 (vmp1) is a potential marker of poor prognosis in her2 positive breast cancer}.
\bjtitle{PLoS One}
\bvolume{14}(\bissue{8}),
\bfpage{0221413}
(\byear{2019})
\end{barticle}
\endbibitem

\bibitem[\protect\citeauthoryear{Raap et~al.}{2018}]{raap2018lobular}
\begin{barticle}
\bauthor{\bsnm{Raap}, \binits{M.}},
\bauthor{\bsnm{Gronewold}, \binits{M.}},
\bauthor{\bsnm{Christgen}, \binits{H.}},
\bauthor{\bsnm{Glage}, \binits{S.}},
\bauthor{\bsnm{Bentires-Alj}, \binits{M.}},
\bauthor{\bsnm{Koren}, \binits{S.}},
\bauthor{\bsnm{Derksen}, \binits{P.W.}},
\bauthor{\bsnm{Boelens}, \binits{M.}},
\bauthor{\bsnm{Jonkers}, \binits{J.}},
\bauthor{\bsnm{Lehmann}, \binits{U.}}, \betal:
\batitle{Lobular carcinoma in situ and invasive lobular breast cancer are characterized by enhanced expression of transcription factor ap-2$\beta$}.
\bjtitle{Laboratory Investigation}
\bvolume{98}(\bissue{1}),
\bfpage{117}--\blpage{129}
(\byear{2018})
\end{barticle}
\endbibitem

\bibitem[\protect\citeauthoryear{Byrne et~al.}{1995}]{byrne1995screening}
\begin{barticle}
\bauthor{\bsnm{Byrne}, \binits{J.A.}},
\bauthor{\bsnm{Tomasetto}, \binits{C.}},
\bauthor{\bsnm{Garnier}, \binits{J.-M.}},
\bauthor{\bsnm{Rouyer}, \binits{N.}},
\bauthor{\bsnm{Mattei}, \binits{M.-G.}},
\bauthor{\bsnm{Bellocq}, \binits{J.-P.}},
\bauthor{\bsnm{Rio}, \binits{M.-C.}},
\bauthor{\bsnm{Basset}, \binits{P.}}:
\batitle{A screening method to identify genes commonly overexpressed in carcinomas and the identification of a novel complementary dna sequence}.
\bjtitle{Cancer Research}
\bvolume{55}(\bissue{13}),
\bfpage{2896}--\blpage{2903}
(\byear{1995})
\end{barticle}
\endbibitem

\bibitem[\protect\citeauthoryear{Balleine et~al.}{2000}]{balleine2000hd52}
\begin{barticle}
\bauthor{\bsnm{Balleine}, \binits{R.L.}},
\bauthor{\bsnm{Fejzo}, \binits{M.S.}},
\bauthor{\bsnm{Sathasivam}, \binits{P.}},
\bauthor{\bsnm{Basset}, \binits{P.}},
\bauthor{\bsnm{Clarke}, \binits{C.L.}},
\bauthor{\bsnm{Byrne}, \binits{J.A.}}:
\batitle{The hd52 (tpd52) gene is a candidate target gene for events resulting in increased 8q21 copy number in human breast carcinoma}.
\bjtitle{Genes, Chromosomes and Cancer}
\bvolume{29}(\bissue{1}),
\bfpage{48}--\blpage{57}
(\byear{2000})
\end{barticle}
\endbibitem

\bibitem[\protect\citeauthoryear{Xie et~al.}{2019}]{xie2019expression}
\begin{barticle}
\bauthor{\bsnm{Xie}, \binits{W.}},
\bauthor{\bsnm{Zhang}, \binits{H.}},
\bauthor{\bsnm{Qin}, \binits{S.}},
\bauthor{\bsnm{Zhang}, \binits{J.}},
\bauthor{\bsnm{Fan}, \binits{X.}},
\bauthor{\bsnm{Yin}, \binits{Y.}},
\bauthor{\bsnm{Liang}, \binits{R.}},
\bauthor{\bsnm{Long}, \binits{H.}},
\bauthor{\bsnm{Yi}, \binits{W.}},
\bauthor{\bsnm{Fu}, \binits{D.}}, \betal:
\batitle{The expression and clinical significance of secretory leukocyte proteinase inhibitor (slpi) in mammary carcinoma using bioinformatics analysis}.
\bjtitle{Gene}
\bvolume{720},
\bfpage{144088}
(\byear{2019})
\end{barticle}
\endbibitem

\bibitem[\protect\citeauthoryear{Nomair et~al.}{2021}]{nomair2021scgb3a1}
\begin{barticle}
\bauthor{\bsnm{Nomair}, \binits{A.M.}},
\bauthor{\bsnm{Ahmed}, \binits{S.S.}},
\bauthor{\bsnm{Mohammed}, \binits{A.F.}},
\bauthor{\bsnm{El~Mansy}, \binits{H.}},
\bauthor{\bsnm{Nomeir}, \binits{H.M.}}:
\batitle{Scgb3a1 gene dna methylation status is associated with breast cancer in egyptian female patients}.
\bjtitle{Egyptian Journal of Medical Human Genetics}
\bvolume{22},
\bfpage{1}--\blpage{12}
(\byear{2021})
\end{barticle}
\endbibitem

\bibitem[\protect\citeauthoryear{Zhong et~al.}{2021}]{zhong2021low}
\begin{barticle}
\bauthor{\bsnm{Zhong}, \binits{P.}},
\bauthor{\bsnm{Shu}, \binits{R.}},
\bauthor{\bsnm{Wu}, \binits{H.}},
\bauthor{\bsnm{Liu}, \binits{Z.}},
\bauthor{\bsnm{Shen}, \binits{X.}},
\bauthor{\bsnm{Hu}, \binits{Y.}}:
\batitle{Low krt15 expression is associated with poor prognosis in patients with breast invasive carcinoma}.
\bjtitle{Experimental and Therapeutic Medicine}
\bvolume{21}(\bissue{4}),
\bfpage{1}--\blpage{1}
(\byear{2021})
\end{barticle}
\endbibitem

\bibitem[\protect\citeauthoryear{Sirois et~al.}{2019}]{sirois2019unique}
\begin{barticle}
\bauthor{\bsnm{Sirois}, \binits{I.}},
\bauthor{\bsnm{Aguilar-Mahecha}, \binits{A.}},
\bauthor{\bsnm{Lafleur}, \binits{J.}},
\bauthor{\bsnm{Fowler}, \binits{E.}},
\bauthor{\bsnm{Vu}, \binits{V.}},
\bauthor{\bsnm{Scriver}, \binits{M.}},
\bauthor{\bsnm{Buchanan}, \binits{M.}},
\bauthor{\bsnm{Chabot}, \binits{C.}},
\bauthor{\bsnm{Ramanathan}, \binits{A.}},
\bauthor{\bsnm{Balachandran}, \binits{B.}}, \betal:
\batitle{A unique morphological phenotype in chemoresistant triple-negative breast cancer reveals metabolic reprogramming and plin4 expression as a molecular vulnerability}.
\bjtitle{Molecular Cancer Research}
\bvolume{17}(\bissue{12}),
\bfpage{2492}--\blpage{2507}
(\byear{2019})
\end{barticle}
\endbibitem

\bibitem[\protect\citeauthoryear{Gopalakrishnan et~al.}{2015}]{gopalakrishnan2015cranial}
\begin{barticle}
\bauthor{\bsnm{Gopalakrishnan}, \binits{S.}},
\bauthor{\bsnm{Comai}, \binits{G.}},
\bauthor{\bsnm{Sambasivan}, \binits{R.}},
\bauthor{\bsnm{Francou}, \binits{A.}},
\bauthor{\bsnm{Kelly}, \binits{R.G.}},
\bauthor{\bsnm{Tajbakhsh}, \binits{S.}}:
\batitle{A cranial mesoderm origin for esophagus striated muscles}.
\bjtitle{Developmental cell}
\bvolume{34}(\bissue{6}),
\bfpage{694}--\blpage{704}
(\byear{2015})
\end{barticle}
\endbibitem

\bibitem[\protect\citeauthoryear{Reardon et~al.}{2012}]{reardon2012cdh1}
\begin{barticle}
\bauthor{\bsnm{Reardon}, \binits{S.N.}},
\bauthor{\bsnm{King}, \binits{M.L.}},
\bauthor{\bsnm{MacLean}, \binits{J.A.}},
\bauthor{\bsnm{Mann}, \binits{J.L.}},
\bauthor{\bsnm{DeMayo}, \binits{F.J.}},
\bauthor{\bsnm{Lydon}, \binits{J.P.}},
\bauthor{\bsnm{Hayashi}, \binits{K.}}:
\batitle{Cdh1 is essential for endometrial differentiation, gland development, and adult function in the mouse uterus}.
\bjtitle{Biology of Reproduction}
\bvolume{86}(\bissue{5}),
\bfpage{141}--\blpage{1}
(\byear{2012})
\end{barticle}
\endbibitem

\bibitem[\protect\citeauthoryear{Kobayashi and Yamamoto}{2007}]{kobayashi2007regulation}
\begin{barticle}
\bauthor{\bsnm{Kobayashi}, \binits{M.}},
\bauthor{\bsnm{Yamamoto}, \binits{M.}}:
\batitle{Regulation of gata1 gene expression}.
\bjtitle{Journal of Biochemistry}
\bvolume{142}(\bissue{1}),
\bfpage{1}--\blpage{10}
(\byear{2007})
\end{barticle}
\endbibitem

\bibitem[\protect\citeauthoryear{Isaac et~al.}{1998}]{isaac1998tbx}
\begin{barticle}
\bauthor{\bsnm{Isaac}, \binits{A.}},
\bauthor{\bsnm{Rodriguez-Esteban}, \binits{C.}},
\bauthor{\bsnm{Ryan}, \binits{A.}},
\bauthor{\bsnm{Altabef}, \binits{M.}},
\bauthor{\bsnm{Tsukui}, \binits{T.}},
\bauthor{\bsnm{Patel}, \binits{K.}},
\bauthor{\bsnm{Tickle}, \binits{C.}},
\bauthor{\bsnm{Izpis{\'u}a-Belmonte}, \binits{J.-C.}}:
\batitle{Tbx genes and limb identity in chick embryo development}.
\bjtitle{Development}
\bvolume{125}(\bissue{10}),
\bfpage{1867}--\blpage{1875}
(\byear{1998})
\end{barticle}
\endbibitem

\bibitem[\protect\citeauthoryear{Porter et~al.}{1997}]{porter1997lhx2}
\begin{barticle}
\bauthor{\bsnm{Porter}, \binits{F.D.}},
\bauthor{\bsnm{Drago}, \binits{J.}},
\bauthor{\bsnm{Xu}, \binits{Y.}},
\bauthor{\bsnm{Cheema}, \binits{S.S.}},
\bauthor{\bsnm{Wassif}, \binits{C.}},
\bauthor{\bsnm{Huang}, \binits{S.-P.}},
\bauthor{\bsnm{Lee}, \binits{E.}},
\bauthor{\bsnm{Grinberg}, \binits{A.}},
\bauthor{\bsnm{Massalas}, \binits{J.S.}},
\bauthor{\bsnm{Bodine}, \binits{D.}}, \betal:
\batitle{Lhx2, a lim homeobox gene, is required for eye, forebrain, and definitive erythrocyte development}.
\bjtitle{Development}
\bvolume{124}(\bissue{15}),
\bfpage{2935}--\blpage{2944}
(\byear{1997})
\end{barticle}
\endbibitem

\bibitem[\protect\citeauthoryear{Xu et~al.}{2010}]{xu2010pamm}
\begin{barticle}
\bauthor{\bsnm{Xu}, \binits{Y.}},
\bauthor{\bsnm{Morse}, \binits{L.R.}},
\bauthor{\bsnm{Da~Silva}, \binits{R.A.B.}},
\bauthor{\bsnm{Odgren}, \binits{P.R.}},
\bauthor{\bsnm{Sasaki}, \binits{H.}},
\bauthor{\bsnm{Stashenko}, \binits{P.}},
\bauthor{\bsnm{Battaglino}, \binits{R.A.}}:
\batitle{Pamm: a redox regulatory protein that modulates osteoclast differentiation}.
\bjtitle{Antioxidants \& redox signaling}
\bvolume{13}(\bissue{1}),
\bfpage{27}--\blpage{37}
(\byear{2010})
\end{barticle}
\endbibitem

\bibitem[\protect\citeauthoryear{O'Shaughnessy}{2019}]{o2019targeting}
\begin{barticle}
\bauthor{\bsnm{O'Shaughnessy}, \binits{R.}}:
\batitle{Targeting tryptophan transport and breakdown in basal cell carcinoma}.
\bjtitle{British Journal of Dermatology}
\bvolume{180}(\bissue{1}),
\bfpage{16}--\blpage{17}
(\byear{2019})
\end{barticle}
\endbibitem

\bibitem[\protect\citeauthoryear{Luan et~al.}{2021}]{luan2021identification}
\begin{botherref}
\oauthor{\bsnm{Luan}, \binits{H.}},
\oauthor{\bsnm{He}, \binits{Y.}},
\oauthor{\bsnm{Jian}, \binits{L.}},
\oauthor{\bsnm{Zhang}, \binits{T.}},
\oauthor{\bsnm{Zhou}, \binits{L.}}:
Identification of metastasis-associated gene and its correlation with immune infiltrates for skin cutaneous melanoma
(2021)
\end{botherref}
\endbibitem

\bibitem[\protect\citeauthoryear{He et~al.}{2023}]{he2023hopx}
\begin{barticle}
\bauthor{\bsnm{He}, \binits{S.}},
\bauthor{\bsnm{Ding}, \binits{Y.}},
\bauthor{\bsnm{Ji}, \binits{Z.}},
\bauthor{\bsnm{Yuan}, \binits{B.}},
\bauthor{\bsnm{Chen}, \binits{J.}},
\bauthor{\bsnm{Ren}, \binits{W.}}:
\batitle{Hopx is a tumor-suppressive biomarker that corresponds to t cell infiltration in skin cutaneous melanoma}.
\bjtitle{Cancer Cell International}
\bvolume{23}(\bissue{1}),
\bfpage{122}
(\byear{2023})
\end{barticle}
\endbibitem

\bibitem[\protect\citeauthoryear{Halifu et~al.}{2016}]{halifu2016wnt1}
\begin{barticle}
\bauthor{\bsnm{Halifu}, \binits{Y.}},
\bauthor{\bsnm{Liang}, \binits{J.}},
\bauthor{\bsnm{Zeng}, \binits{X.}},
\bauthor{\bsnm{Ding}, \binits{Y.}},
\bauthor{\bsnm{Zhang}, \binits{X.}},
\bauthor{\bsnm{Jin}, \binits{T.}},
\bauthor{\bsnm{Yakeya}, \binits{B.}},
\bauthor{\bsnm{Abudu}, \binits{D.}},
\bauthor{\bsnm{Zhou}, \binits{Y.}},
\bauthor{\bsnm{Liu}, \binits{X.}}, \betal:
\batitle{Wnt1 and sfrp1 as potential prognostic factors and therapeutic targets in cutaneous squamous cell carcinoma}.
\bjtitle{Genet Mol Res}
\bvolume{15}(\bissue{2}),
\bfpage{8187}
(\byear{2016})
\end{barticle}
\endbibitem

\bibitem[\protect\citeauthoryear{Chen et~al.}{2021}]{chen2021seven}
\begin{barticle}
\bauthor{\bsnm{Chen}, \binits{H.}},
\bauthor{\bsnm{Yang}, \binits{J.}},
\bauthor{\bsnm{Wu}, \binits{W.}}:
\batitle{Seven key hub genes identified by gene co-expression network in cutaneous squamous cell carcinoma}.
\bjtitle{BMC Cancer}
\bvolume{21},
\bfpage{1}--\blpage{12}
(\byear{2021})
\end{barticle}
\endbibitem

\end{thebibliography}

\end{document}